\documentclass[useAMS,usenatbib]{mn2e}
\usepackage[dvips]{graphicx}
\usepackage{amssymb}

\title[Light curve studies of nearby Type Ia Supernovae\phantom{0}\\with a Multi-band Stretch method]{Light curve studies of nearby Type Ia Supernovae \\with a Multi-band Stretch method}
\author[N. Takanashi, M. Doi, and, N. Yasuda]{N. Takanashi$^{1,2}$\thanks{E-mail:
naohiro.takanashi@nao.ac.jp}, M. Doi$^{1}$ and N. Yasuda$^{3}$\\
$^{1}$Institute of Astronomy, Graduate School of Science, University of Tokyo, 2-21-1 Osawa, Mitaka, Tokyo 181-0015, Japan.\\
$^{2}$National Astronomical Observatory of Japan, Mitaka 181-8588, Japan.\\
$^{3}$Institute for Cosmic Ray Research, University of Tokyo, Kashiwa 277-8582, Japan.}
\begin{document}

%\date{Accepted 1988 December 15. Received 1988 December 14; in original form 1988 October 11}

\pagerange{\pageref{firstpage}--\pageref{lastpage}} \pubyear{2007}

\maketitle

\label{firstpage}

\begin{abstract}
We create new U, B, V, R and I-band light curve templates of type Ia Supernovae (SNe Ia) and re-analyze 122 nearby (redshift $<$ 0.11) SNe Ia using a new ``Multi-band Stretch method,'' which is a revised Stretch method (cf. \citealt{per97,gol01}) extended to five bands. We find (i) our I-band template can fit about 90\% of SNe Ia I-band light curves, (ii) relationships between luminosity, colours and stretch factors, (iii) possible sub-groups of SNe Ia, and (iv) the ratio of total to selective extinction $R$ in other galaxies can be consistent with that in the Milky Way under the assumption that SNe Ia have diversity in their intrinsic colour. Based on these results, we discuss how to select subsets of SNe Ia to serve as good distance indicators for cosmology. We find two possibilities: one is to choose ``BV bluest'' ($-0.14 < (B-V)_{max} \le -0.10$) objects and the other is to use only SNe Ia which occur in E or S0 galaxies. Within these subsets, we find the root mean square (r.m.s.) of peak B-band magnitudes is 0.17 mag (``BV bluest'' sample) and 0.12 mag (E or S0 sample).
\end{abstract}


\section{INTRODUCTION}

SNe Ia are some of the most important objects in observational cosmology because of their role as distance indicators. SNe Ia are bright enough to be observed at high redshifts, they have uniform peak luminosity (cf.\citealt{bra92}), and they are expected to evolve less from low z to high z than other objects such as galaxies (cf. \citealt{rie99b}). Many studies have used SN Ia as distance indicators to show the accelerating expansion of the universe (e.g \citealt{rie98,per99,kno03,rie04,ast06,woo07}). All of these works are based on the empirical relationship between light curve shapes and peak luminosity (cf. \citealt{phi93}). From the 1990s onward, as more low-redshift SNe Ia have been found, the relation has been revised (cf. \citealt{ham96,rie99a,alt04,rei05}).

Recently, the number of well-measured SNe Ia has grown larger and larger, especially at intermediate and high redshifts. There are many supernova survey programs currently running; for example, the Lick Observatory and Tenagra Observatory Supernova Searches (LOTOSS\footnote{http://astro.berkeley.edu/\~{}bait/lotoss.html}), the Nearby Supernova Factory (SNfactory, \citealt{ald02}), the SDSS-II Supernova Survey (SDSS SN Survey, \citealt{fri08,sak08}), Equation of State: SupErNovae trace Cosmic Expansion (ESSENCE, \citealt{woo07}), and the SuperNova Legacy Survey (SNLS, \citealt{pri05}). Most of SNe Ia found in these surveys are observed in multiple passbands and over many epochs. These high quality samples have given us clues to understand the nature of SN Ia. For example, \cite{ham96c} showed that brighter SNe Ia, like SN1991T, tend to occur in galaxies with active star formation, while fainter SNe Ia, such as SN1991bg, tend to occur in galaxies which lack star formation (see also \citealt{how01,van05,sul06}). \cite{ben05} discussed the intrinsic diversity of the spectroscopically normal SN Ia (called "Branch normal", \citealt{bra93}). \cite{qui07} pointed out there are two subgroups within Branch normal SN Ia. Unusual SNe Ia have also been found: SN2000cx and SN2002cx were reported as an extreme sample of spectroscopically peculiar SN Ia \citep{li01,li03}.; \cite{how06} reported SNLS-03D3bb was a Super-Chandrasekhar-mass supernova with an exceptionally high luminosity and low kinetic energy. H$\alpha$ emission in the spectra of SN2002ic and SN2005gj at early phases indicates the existence of circumsteller material around the progenitor (\citealt{den04,ald06} and so on).

Although SNe Ia seem to be good standard candles for cosmological studies, there are several issues we must understand in order to use them optimally. The main issues are (1) intrinsic photometric and spectroscopic diversity of SNe Ia and (2) extinction in host galaxies. Both issues severely limit the use of SNe Ia as a standard candles if we can not deal with them properly. Recent studies have revealed an intrinsic diversity of SNe Ia; for example, \cite{man05} and \cite{sul06} showed there are two types of SN Ia, called ``prompt'' and ``delayed'', which differ in the distribution of their stretch factors, and \cite{ber06} demonstrated how spectroscopic diversity can affect significantly the intrinsic colours of SNe Ia and K-corrections. The latter issue is important if we want to use larger SN Ia samples. Correction of extinction by dust in the host galaxy is necessary because almost all SNe Ia may be affected by the dust. Previous works insisted that the ratio of total to selective extinction, $R_{B}$, in other galaxies is smaller than that of the Milky Way; see \cite{alt04} (hereafter ALT04), \cite{phi99}, and \cite{kno03}.

In this paper, we discuss these issues using a sample of nearby SNe Ia. We re-analyze published data on nearby SNe Ia with an improved ``Multi-band Stretch method'' which is based on the usual ``stretch method'' (\citealt{per97,gol01}).

\section{PHOTOMETRIC DATA}

\begin{table}
\begin{center}
\begin{minipage}{25em}
\caption{Number of SNe Ia in each band}
\label{sample}
\end{minipage}
\end{center}
\begin{center}
\small
\begin{tabular}{cccccc} \hline\hline
Sample & U-band & B-band & V-band & R-band & I-band \\ \hline
1A & 49 & 122 & 122 & 101 & 115 \\
1B & 44 & 108 & 108 & \phantom{0}88 & 102 \\ \hline
\end{tabular}\\
\vspace*{0.6em}
{\footnotesize
\begin{minipage}{26em}
1A are all SNe Ia we analyzed in this paper.\\
1B are selected SNe Ia based on $\chi^2$ (see \S3.2).
\end{minipage}}
\end{center}
\end{table} %

We use U, B, V, R and I-band light curves in Vega system of 122 SNe in this work. All of the sample have B and V-band photometry, but they don't always have U, R and I-band photometry (see Table \ref{sample}). All photometric data are obtained from published papers, mainly from \cite{jha06} (hereafter JHA06), \cite{rie99a} and \cite{ham96}. We also took photometric data from \cite{for93}, \cite{wel94}, \cite{ric95}, \cite{pat96}, \cite{lir98}, \cite{sun99}, \cite{kri00}, \cite{mod01}, \cite{ho01}, \cite{kri01}, \cite{str02}, \cite{kri03}, \cite{li03}, \cite{vin03}, \cite{val03}, \cite{vin03}, \cite{gar04}, \cite{ger04}, \cite{kri04a}, \cite{kri04b}, \cite{pig04}, \cite{anu05}, \cite{kun05} and \cite{sah06}. All of SNe Ia used in this paper have B and V-band data. We analyzed only bands in which there are more than 3 measurements. For example, if a SN had five B-band measurements, five V-band measurements and two I-band measurements, we did not include its I-band photometry in our analysis. There are some SNe Ia which were observed by different groups independently, but we didn't combine these light curves to avoid systematic errors; instead, we used the light curve with the largest number of epochs. Many of our data are also found in the compilations of \cite{alt04} and \cite{rei05}. Figure \ref{z_count} shows the redshift distribution of our sample. 

According to \cite{rei05}, there are five ``SN1991T-like'' SNe (SN1991T, SN1995ac, SN1995bd, SN1997br, SN2000cx) and seven ``SN1991bg-like'' SNe (SN1991bg, SN1992K, SN1997cn, SN1998bp, SN1998de, SN1999by, SN1999da) in our sample. The other 111 SNe are either ``Branch normal'' or lack detailed spectroscopic analysis.

\section{ANALYSIS}

\subsection{Light Curve Parameterization}

There are several ways to describe the shape of a supernova's light curve, but we can classify them roughly into two types. One type concentrates on the data in a single passband, such as the ``$\Delta$ m15'' \citep{phi93} or ``Stretch'' \citep{per97} algorithms. The other type includes measurements of light curves in multiple passbands, such as the ``MLCS, MLCS2k2'' \citep{rie96,jha07}, ``CMAGIC'' \citep{wan03}, and ``SALT, SALT2'' \citep{guy05,guy07} algorithms. The latter type is more complex and employs SN colours as a part of the analysis. \cite{pri06} proposed a new method, a combination of $\Delta$ m15 and MLCS, to gain the advantages from both approaches.

For this work, we chose the single-band stretch method as a base, since it is simple to parameterize light curves: we just need one template for each band. We created new U, B, V, R and I-band templates (see \S3.4), which we simply stretch to fit observed light curves, without having to account for additional assumptions such as dust extinction. This method works well when we simply want to assign a light curve shape and peak apparent brightness in each band.

Usually, the stretch method has been applied only to B  or V-band light curves. However, in this work, we apply the stretch method to data in all passsbands: U, B, V, R and I. In other words, we fit up to 11 parameters to each event: U, B, V, R and I-band stretch factors, U, B, V, R and I-band peak magnitudes, and a time $t_{Bmax}$ for maximum light in the B-band. The reason that we apply a stretch factor to U, B, V, R and I-band independently instead of a common stretch factor like SALT \citep{guy05} is that we wish to investigate the diversity of light curve shapes between passbands. We call this method the "Multi-band Stretch method".

It is well known that there are some peculiarities in redder band light curve shapes, especially in I-band. There are usually two peaks in I-band light curves, but some SNe Ia lack the second peak. There are very few of these SNe Ia without the second peak in our sample, so we use only ``normal'' I-band template for our analysis (see \S4.3).

Using information in multiple passbands improves the determination of the time of B-band maximum brightness. By using data from the second peak in R and I-band around 30-40 days after B-band maximum, we can determine the time of B-band maximum even if there are no observations around the first peak. In addition, as a result of fitting, we can use colour information to study the intrinsic colours of SNe Ia, the effects of dust extinction, and so on.

Please note that we use the inverse of the usual stretch factor ($1/s$) for easier comparison with the $\Delta m 15$ parameter.

\subsection{Fitting Algorithm}

We used a reduced $\chi^2$ method for fitting templates to observations. Our definition of reduced $\chi^2$ is as follows;

\begin{equation}
\chi^{2} = \frac{1}{DOF} \times \sum_{X}\sum^{n}_{i=0}\Bigg(\frac{\Delta m}{\sqrt{\sigma^2_{obs} + \sigma^2_{temp}}}\Bigg)^2\\
\end{equation}
\begin{eqnarray}
(\Delta m = m_{Xi} - K_{X,(t_i-t_{Bmax})/(1+z)} - \phantom{000000000000000000000000000000000000000}\nonumber \\ 
M^{temp}_X \{(1+z) {\times} (1 / s_{(X)}) \times (t_i-t_{Bmax})\} - X_{max} - \mu) \phantom{00000000000000000000000}\nonumber
\end{eqnarray}

Here, $n$ is the total number of observed epochs, $DOF$ is the number of degrees of freedom in the fitting, $m_{Xi}$ is the observed apparent magnitude in X-band after correction for Galactic dust extinction from \cite{sch98}, $t_{Bmax}$ is time of the maximum in B-band, $K_{X,t}$ is the K-correction in the X-band\footnote{We didn't need cross filter K-corrections because the typical redshift of our sample is so small.} magnitude at time $t-t_{Bmax}$, $\mu$ is the distance modulus derived from a recession velocity, $M^{temp}_{X}(t)$ is the template's magnitude at time t,  $s_{(X)}$ is the stretch factor in X-band, $X_{max}$ is the peak magnitude in X-band at the time of B-band maximum, $\sigma_{obs}$ is the photometric error at the measurement, and $\sigma_{template}$ is the error of template at the epoch. We fit the parameters $t_{Bmax}, s_{(X)}, X_{max}$ to each set of curves of one supernova.Although our Eq.(1) is expressed in magnitudes, we performed the fitting in linear scale. U, B, V, R and I-band light curves were treated with equal weights. We used the recession velocity of each host galaxy and cosmological parameters ($H_{0} = 70.8 {\rm \thinspace km/s/Mpc}, \thinspace \Omega_{M} = 0.262, \thinspace \Omega_{\Lambda} = 0.738$ from \citealt{spe07}) to determine the distance modulus. Note that we did not introduce any colour terms in Eq.(1). We estimated the typical size of errors in our fitting procedure with Monte Carlo simulations. We fitted 30 artificial multi-band light curves which made from light curve templates and obtained the dispersion for each fitted parameter.

In our Multi-band Stretch method, the template can be applied to the observed light curve anywhere from -10 days to +80 days after B-band maximum (-10 days to +70 days for U-band only). These ranges are wider than those of previous papers. For example, \cite{gol01} used a range of -25 days to +50 days. As a result, we can use information from later phases, especially around the second peak in R and I-band.

\subsection{K-corrections}

Even though SNe Ia are simpler objects than galaxies, their K-corrections are not so simple since one must account for their spectral evolution. In this work, we define a K-correction as follows: 

\begin{equation} 
K^{counts}_{X,t} = 2.5 \log{(1+z)} + 2.5 \log{\frac{\int{\lambda}F(\lambda,t)S_{X}(\lambda)d\lambda}{\int{\lambda}F[\lambda/(1+z),t]S_{X}(\lambda)d\lambda}}
\end{equation} 

Here $F(\lambda,t)$ is the spectral energy distribution (SED) of the SN at time $t$, $S_X(\lambda)$ is the effective spectral transmission in the X-band, and $K_{X,t}$ is the value of K-correction at time $t$. See \cite{kim96} and \cite{nug02} for details. We used Bessell U, B, V, R and I-band filter shapes (c.f. \citealt{bes90}) for the calculation.

Studies of K-corrections in nearby SN Ia samples goes back to \cite{ham93}. \cite{kim96} followed this work, and \cite{nug02} provided a synthetic spectral library for SNe Ia. We used the Nugent's model version1.1\footnote{http://supernova.lbl.gov/\~{}nugent/nugent\_templates.html} for our calculations of K-corrections. There are three type of spectral models, named ``Branch normal'', ``SN1991bg-like'', and ``SN1991T-like'' \citep{bra93}. We use the classification of each SN in \cite{rei05} to choose a spectral model, or used the ``Branch normal'' spectral model when 
there was no spectral classification..

\subsection{Template}

Figure \ref{templates} shows our new U, B, V, R and I-band templates of SN Ia light curves. The templates were composed from 110 nearby ($z < 0.05$, 90\% have $z < 0.03$) SNe Ia light curves. The recipe to make the templates is as follows: 

\begin{enumerate} 
\item Select a densely-observed "seed" light curve (we selected SN2001el \citep{kri03} for this work) and interpolate epochs.  
\item Stretch the other 110 nearby SNe Ia light curves to fit in the current templates without K-corrections, then take the weighted average of all.  
\item Iterate step (ii) until the templates become smooth.  
\item Finally, stretch each template to fit in the $s=1$ light curve defined by \cite{per97} template\footnote{http://supernova.lbl.gov/\~{}nugent/nugent\_templates.html}.

\end{enumerate} 

We didn't apply K-corrections to make the light curve template because well-observed SNe Ia which control the shape of light curve template are at the lowest redshift, and the size of K-correction is smaller than typical photometric errors in all bands.

We calculated the residuals of each light curve from the template (see Figure \ref{templates}). We regard this scatter from the template as a result of the intrinsic diversity of SNe. The templates are available from our web site\footnote{http://www.ioa.s.u-tokyo.ac.jp/\~{}takanashi/works/SN/template/}.

\begin{table*}
\rotatebox{90}{% 90"x‰ñ"]'³'¹'é
\begin{minipage}{\textheight}
\caption{The Best Fitting Parameters of 108 Selected SNe Ia} \label{data}
\centering
{\small
\tabcolsep = 1mm
\begin{tabular}{lcccccccccccccl} \hline\hline
SN name & U & $s_{(U)}$ & B & $s_{(B)}$ & V & $s_{(V)}$ & R & $s_{(R)}$ & I & $s_{(I)}$ & $\mu$ & $\frac{\chi^2}{d.o.f.}$ & host & ref. \\\hline
SN1989B & -18.61(0.03) & 0.988(0.018) & -18.69(0.04) & 0.902(0.021) & -18.98(0.02) & 0.999(0.012) & -19.20(0.02) & 0.944(0.020) & -19.26(0.02) & 0.936(0.028) & 30.89& 0.476& - & (1) \\
SN1990O$^{**}$ & - & - & -19.41(0.02) & 1.109(0.032) & -19.30(0.02) & 1.138(0.024) & -19.29(0.03) & 1.104(0.025) & -18.95(0.02) & 1.061(0.026) & 35.60& 0.891& - & (2) \\
SN1990T & - & - & -18.92(0.06) & 1.067(0.048) & -19.06(0.06) & 0.995(0.032) & -18.95(0.03) & 1.050(0.020) & -18.74(0.03) & 1.006(0.017) & 36.18& 0.387& E & (2) \\
SN1990Y & - & - & -18.29(0.07) & 1.076(0.054) & -18.67(0.09) & 1.011(0.043) & -18.86(0.09) & 0.954(0.045) & -18.79(0.17) & 0.864(0.056) & 35.95& 1.445& E & (2) \\
SN1990af$^{*}$ & - & - & -18.91(0.01) & 0.787(0.014) & -18.87(0.01) & 0.789(0.011) & - & - & - & - & 36.69& 0.392& E & (2) \\
SN1991S & - & - & -19.32(0.13) & 0.952(0.067) & -19.02(0.04) & 1.096(0.034) & -19.00(0.03) & 1.091(0.021) & -18.85(0.06) & 0.996(0.039) & 36.89& 0.478& - & (2) \\
SN1991T & -21.26(0.02) & 1.155(0.023) & -20.77(0.04) & 1.119(0.025) & -20.92(0.02) & 1.102(0.002) & -20.93(0.03) & 1.092(0.030) & -20.73(0.04) & 1.103(0.029) & 32.34& 2.363& - & (3) \\
SN1991U$^{**}$ & - & - & -19.26(0.06) & 1.005(0.028) & -19.14(0.06) & 1.102(0.032) & -19.17(0.03) & 1.117(0.025) & -19.12(0.02) & 0.989(0.024) & 35.65& 0.427& - & (2) \\
SN1991ag$^{*}$ & - & - & -19.41(0.02) & 1.120(0.016) & -19.38(0.03) & 1.110(0.014) & -19.38(0.03) & 1.070(0.025) & -19.07(0.01) & 1.065(0.014) & 33.94& 0.311& - & (2) \\
SN1992J & - & - & -18.69(0.07) & 0.824(0.037) & -18.89(0.21) & 0.784(0.111) & - & - & -18.55(0.03) & 0.804(0.024) & 36.44& 0.543& - & (2) \\
SN1992K & - & - & -18.39(0.07) & 0.590(0.036) & -18.60(0.23) & 0.636(0.101) & - & - & -18.17(0.04) & 0.801(0.022) & 33.37& 0.831& - & (2) \\
SN1992P$^{**}$ & - & - & -19.22(0.02) & 1.035(0.020) & -19.11(0.01) & 1.044(0.016) & - & - & -18.75(0.03) & 1.127(0.057) & 35.27& 1.358& - & (2) \\
SN1992ae$^{*}$ & - & - & -19.09(0.06) & 0.949(0.027) & -19.03(0.02) & 0.942(0.020) & - & - & - & - & 37.58& 0.390& E & (2) \\
SN1992al$^{**}$ & - & - & -19.42(0.01) & 0.984(0.014) & -19.28(0.01) & 1.030(0.013) & -19.23(0.02) & 1.002(0.011) & -18.93(0.01) & 1.024(0.009) & 33.88& 0.292& - & (2) \\
SN1992aq$^{**}$ & - & - & -18.95(0.03) & 0.840(0.034) & -18.82(0.02) & 0.901(0.028) & - & - & -18.38(0.07) & 0.818(0.050) & 38.25& 0.425& - & (2) \\
SN1992au$^{*}$ & - & - & -18.98(0.20) & 0.785(0.061) & -18.91(0.19) & 0.844(0.134) & - & - & -18.50(0.03) & 0.798(0.025) & 37.15& 0.487& E & (2) \\
SN1992bc$^{**}$ & - & - & -19.60(0.03) & 1.081(0.015) & -19.47(0.01) & 1.039(0.034) & -19.41(0.03) & 1.012(0.008) & -18.99(0.04) & 1.027(0.017) & 34.66& 2.307& - & (2) \\
SN1992bg$^{**}$ & - & - & -19.21(0.04) & 0.997(0.033) & -19.10(0.02) & 1.025(0.014) & - & - & -18.78(0.03) & 0.981(0.013) & 35.89& 0.677& - & (2) \\
SN1992bh & - & - & -18.84(0.02) & 1.024(0.022) & -18.82(0.02) & 1.044(0.022) & - & - & -18.58(0.02) & 1.028(0.020) & 36.44& 0.402& - & (2) \\
SN1992bk$^{*}$ & - & - & -18.89(0.04) & 0.796(0.029) & -18.83(0.05) & 0.806(0.025) & - & - & -18.64(0.08) & 0.740(0.045) & 37.00& 0.941& E & (2) \\
SN1992bl$^{*}$ & - & - & -19.03(0.03) & 0.806(0.020) & -18.94(0.02) & 0.836(0.013) & - & - & -18.62(0.02) & 0.806(0.013) & 36.34& 1.117& E & (2) \\
SN1992bo$^{*}$ & - & - & -18.74(0.02) & 0.773(0.011) & -18.69(0.02) & 0.829(0.011) & -18.75(0.01) & 0.781(0.006) & -18.52(0.01) & 0.780(0.005) & 34.49& 0.761& E & (2) \\
SN1992bp & - & - & -19.43(0.02) & 0.846(0.028) & -19.28(0.01) & 0.912(0.018) & - & - & -18.81(0.03) & 0.960(0.023) & 37.69& 1.547& E & (2) \\
SN1992br$^{*}$ & - & - & -18.56(0.07) & 0.636(0.021) & -18.46(0.03) & 0.741(0.028) & - & - & - & - & 37.94& 0.709& E & (2) \\
SN1992bs$^{**}$ & - & - & -18.94(0.04) & 0.996(0.019) & -18.82(0.03) & 1.015(0.015) & - & - & - & - & 37.19& 0.375& - & (2) \\
SN1993B$^{**}$ & - & - & -19.03(0.03) & 0.930(0.023) & -18.89(0.01) & 1.071(0.019) & - & - & -18.59(0.03) & 1.018(0.029) & 37.39& 0.882& - & (2) \\
SN1993H & - & - & -18.38(0.02) & 0.753(0.013) & -18.52(0.02) & 0.793(0.007) & -18.59(0.01) & 0.751(0.011) & -18.52(0.03) & 0.719(0.012) & 35.08& 0.938& - & (2) \\
SN1993L & - & - & -17.74(0.07) & 1.144(0.049) & -18.21(0.03) & 1.009(0.018) & -18.34(0.04) & 0.980(0.019) & -17.75(0.34) & 1.138(0.149) & 31.47& 1.511& - & (4) \\
SN1993O$^{**}$ & - & - & -19.14(0.02) & 0.955(0.018) & -19.01(0.02) & 0.994(0.018) & - & - & -18.63(0.02) & 0.992(0.029) & 36.76& 0.933& E & (2) \\
SN1993ac$^{*}$ & - & - & -18.95(0.04) & 0.931(0.029) & -18.92(0.07) & 0.909(0.043) & -18.87(0.02) & 1.166(0.038) & -18.63(0.04) & 1.067(0.042) & 36.64& 1.662& E & (5) \\
SN1993ae & - & - & -19.20(0.03) & 0.838(0.022) & -19.23(0.04) & 0.804(0.019) & -19.17(0.04) & 0.797(0.021) & -18.97(0.02) & 0.795(0.015) & 34.55& 0.360& - & (5) \\
SN1993ag & - & - & -18.86(0.02) & 0.972(0.041) & -18.86(0.02) & 0.992(0.025) & - & - & -18.54(0.04) & 0.941(0.016) & 36.68& 0.988& E & (2) \\
SN1993ah & - & - & -19.19(0.09) & 0.849(0.035) & -18.84(0.06) & 1.117(0.045) & - & - & -18.81(0.02) & 0.854(0.078) & 35.50& 1.926& - & (2) \\
SN1994D$^{**}$ & -18.99(0.02) & 0.999(0.009) & -18.49(0.02) & 0.838(0.010) & -18.38(0.01) & 0.874(0.007) & -18.41(0.01) & 0.835(0.008) & -18.07(0.02) & 0.860(0.015) & 30.24& 2.570& E & (6) \\
SN1994M$^{*}$ & - & - & -18.73(0.02) & 0.865(0.017) & -18.68(0.02) & 0.920(0.013) & -18.86(0.02) & 0.827(0.013) & -18.59(0.02) & 0.851(0.010) & 34.98& 0.772& E & (5) \\
SN1994Q & - & - & -19.11(0.07) & 1.103(0.057) & -19.12(0.07) & 1.073(0.041) & -19.13(0.03) & 1.131(0.028) & -18.83(0.02) & 1.023(0.025) & 35.47& 0.250& E & (5) \\
SN1994S$^{*}$ & - & - & -19.39(0.02) & 1.092(0.022) & -19.35(0.02) & 1.045(0.026) & -19.28(0.02) & 0.998(0.019) & -18.95(0.01) & 1.064(0.024) & 34.18& 1.414& - & (5) \\
SN1994ae & - & - & -18.68(0.01) & 1.067(0.015) & -18.69(0.02) & 1.067(0.017) & -18.79(0.01) & 1.043(0.022) & -18.37(0.02) & 1.105(0.023) & 31.76& 1.008& - & (7) \\
SN1995D$^{**}$ & - & - & -19.35(0.02) & 1.081(0.012) & -19.24(0.01) & 1.142(0.013) & -19.24(0.01) & 1.104(0.009) & -18.91(0.01) & 1.128(0.010) & 32.55& 0.361& E & (5) \\
SN1995E & - & - & -16.78(0.02) & 0.992(0.028) & -17.43(0.02) & 1.063(0.019) & -17.95(0.02) & 0.968(0.017) & -18.12(0.02) & 0.968(0.010) & 33.49& 0.390& - & (5) \\\hline
\end{tabular}
}
\flushleft
List of parameters of 108 selected SNe Ia using in this paper. * means "BV bluer" SNe Ia and ** means "BV bluest" SNe Ia (see \S4.2). SNe Ia hosted by E or S0 galaxies are listed as "E". References are as follows: (1) Wells et al.(1994), (2) Hamuy et al.(1996), (3) Krisciunas et al.(2004b), (4) Altavilla et al.(2004), (5) Riess et al.(1999), (6) Richmond et al.(1995), (7) Riess et al.(2005), (8) Jha et al.(2006), (9) Krisciunas et al.(2000), (10) Krisciunas et al.(2001), (11) Valentini et al.(2003), (12) Krisciunas et al.(2004a), (13) Vinko et al.(2003), (14) Krisciunas et al.(2003), (15) VSNET, (16) Kuntal et al.(2005)
\end{minipage}}
\end{table*}

\clearpage

\begin{figure}
\rotatebox{90}{% 90"x‰ñ"]'³'¹'é
\begin{minipage}{\textheight}\centering
{\small
\tabcolsep = 1mm
\begin{tabular}{lcccccccccccccl} \hline\hline
SN name & U & $s_{(U)}$ & B & $s_{(B)}$ & V & $s_{(V)}$ & R & $s_{(R)}$ & I & $s_{(I)}$ & $\mu$ & $\frac{\chi^2}{d.o.f.}$ & host & ref. \\\hline
SN1995ac$^{*}$ & - & - & -19.58(0.01) & 1.072(0.013) & -19.52(0.01) & 1.121(0.013) & -19.55(0.01) & 1.063(0.010) & -19.33(0.01) & 1.133(0.011) & 36.68& 1.524& - & (5) \\
SN1995ak & - & - & -18.85(0.07) & 0.836(0.063) & -18.84(0.03) & 0.927(0.022) & -19.03(0.02) & 0.876(0.011) & -18.78(0.03) & 0.959(0.015) & 34.94& 2.208& - & (5) \\
SN1995al & - & - & -18.74(0.02) & 1.089(0.017) & -18.78(0.02) & 1.117(0.015) & -18.78(0.02) & 1.100(0.014) & -18.54(0.02) & 1.112(0.011) & 32.03& 0.273& - & (5) \\
SN1995bd & - & - & -18.94(0.01) & 1.177(0.011) & -19.21(0.01) & 1.128(0.013) & -19.30(0.01) & 1.157(0.015) & -19.01(0.01) & 1.171(0.007) & 34.17& 2.178& - & (5) \\
SN1996C & - & - & -18.70(0.03) & 1.089(0.013) & -18.72(0.02) & 1.059(0.024) & -18.81(0.02) & 1.023(0.017) & -18.47(0.01) & 1.020(0.007) & 35.32& 0.781& - & (5) \\
SN1996X$^{*}$ & - & - & -19.63(0.01) & 0.900(0.006) & -19.57(0.01) & 0.941(0.007) & -19.56(0.01) & 0.909(0.010) & -19.28(0.01) & 0.981(0.014) & 32.60& 0.385& E & (5) \\
SN1996Z & - & - & -18.48(0.04) & 0.936(0.031) & -18.73(0.02) & 0.994(0.051) & -18.85(0.03) & 0.868(0.072) & - & - & 32.82& 0.245& - & (5) \\
SN1996ab & - & - & -17.48(0.03) & 0.990(0.038) & -17.54(0.03) & 0.849(0.028) & - & - & - & - & 37.08& 1.052& - & (5) \\
SN1996ai & - & - & -14.19(0.01) & 1.108(0.012) & -15.84(0.01) & 1.090(0.011) & -16.62(0.01) & 1.083(0.017) & -17.08(0.01) & 1.053(0.011) & 31.09& 1.494& - & (5) \\
SN1996bk & - & - & -17.41(0.02) & 0.906(0.013) & -17.96(0.01) & 0.778(0.010) & -18.22(0.02) & 0.699(0.007) & -18.31(0.02) & 0.664(0.007) & 32.42& 2.158& E & (5) \\
SN1996bl$^{*}$ & - & - & -19.27(0.02) & 1.005(0.016) & -19.21(0.02) & 1.046(0.018) & -19.22(0.01) & 1.060(0.017) & -18.98(0.01) & 1.032(0.014) & 35.95& 1.014& - & (5) \\
SN1996bv & - & - & -19.01(0.02) & 1.085(0.010) & -19.03(0.01) & 1.180(0.006) & -19.16(0.02) & 1.159(0.028) & -19.05(0.02) & 1.061(0.012) & 34.26& 1.484& - & (5) \\
SN1997E & -19.06(0.02) & 0.906(0.013) & -18.69(0.01) & 0.835(0.007) & -18.70(0.01) & 0.865(0.007) & -18.78(0.01) & 0.816(0.007) & -18.55(0.01) & 0.814(0.005) & 33.78& 0.794& E & (8) \\
SN1997Y$^{**}$ & -19.42(0.05) & 0.938(0.062) & -19.01(0.02) & 0.901(0.012) & -18.88(0.01) & 1.000(0.012) & -18.94(0.01) & 0.971(0.011) & -18.77(0.01) & 0.978(0.009) & 34.25& 0.326& - & (8) \\
SN1997bp & -19.15(0.02) & 1.232(0.022) & -19.11(0.01) & 0.987(0.011) & -19.20(0.01) & 1.166(0.011) & -19.27(0.01) & 1.163(0.012) & -19.02(0.00) & 1.058(0.007) & 33.01& 2.395& - & (8) \\
SN1997bq & -18.53(0.25) & 1.224(0.180) & -18.82(0.09) & 0.819(0.032) & -18.64(0.06) & 1.085(0.077) & -18.78(0.04) & 0.989(0.028) & -18.62(0.02) & 0.949(0.019) & 33.06& 2.994& - & (8) \\
SN1997br & -19.75(0.02) & 0.983(0.009) & -19.37(0.03) & 0.845(0.016) & -19.31(0.01) & 1.157(0.008) & -19.54(0.01) & 1.043(0.007) & -19.48(0.01) & 1.015(0.005) & 32.66&	2.555& - & (8) \\
SN1997cw & -18.75(0.21) & 1.156(0.093) & -18.66(0.03) & 0.967(0.023) & -18.66(0.02) & 1.265(0.012) & -18.94(0.02) & 1.153(0.016) & -18.91(0.01) & 1.064(0.014) & 34.39& 0.830& E & (8) \\
SN1997dg$^{**}$ & -19.56(0.05) & 0.920(0.052) & -19.03(0.02) & 1.017(0.034) & -18.90(0.01) & 1.027(0.016) & -18.99(0.02) & 0.967(0.018) & -18.80(0.02) & 0.908(0.024) & 35.82& 0.695& - & (8) \\
SN1997do$^{*}$ & -19.26(0.05) & 1.061(0.032) & -18.95(0.02) & 1.002(0.011) & -18.89(0.01) & 1.137(0.011) & -18.97(0.01) & 1.093(0.007) & -18.65(0.01) & 1.113(0.006) & 33.25& 1.160& - & (8) \\
SN1997dt & - & - & -16.66(0.02) & 0.971(0.009) & -17.07(0.02) & 1.048(0.016) & -17.41(0.01) & 1.048(0.012) & -17.59(0.01) & 1.015(0.021) & 32.08& 2.996& - & (8) \\
SN1998D & - & - & -18.02(0.14) & 0.904(0.091) & -18.25(0.27) & 0.804(0.194) & -18.50(0.34) & 0.735(0.110) & -17.60(0.06) & 1.105(0.034) & 31.3 & 0.067& - & (8) \\
SN1998V$^{**}$ & -19.73(0.02) & 1.074(0.019) & -19.31(0.01) & 1.012(0.007) & -19.20(0.00) & 1.046(0.005) & -19.25(0.01) & 1.031(0.008) & -19.02(0.01) & 1.026(0.010) & 34.33& 0.877& - & (8) \\
SN1998ab$^{**}$ & -19.81(0.06) & 1.015(0.037) & -19.35(0.02) & 0.958(0.011) & -19.24(0.01) & 1.107(0.006) & -19.36(0.01) & 1.125(0.013) & -19.19(0.01) & 1.054(0.014) & 35.40& 2.314& - & (8) \\
SN1998bu & -19.28(0.02) & 1.147(0.009) & -19.09(0.01) & 1.003(0.005) & -19.36(0.00) & 1.013(0.003) & -19.54(0.01) & 0.984(0.005) & -19.52(0.01) & 1.006(0.004) & 31.20& 0.623& - & (8) \\
SN1998dh & -19.12(0.04) & 1.043(0.021) & -18.71(0.01) & 0.937(0.005) & -18.74(0.01) & 1.006(0.006) & -18.82(0.01) & 0.969(0.008) & -18.61(0.01) & 0.963(0.005) & 32.59& 0.358& - & (8) \\
SN1998dk & -19.27(0.10) & 1.016(0.071) & -18.73(0.03) & 0.955(0.043) & -18.80(0.07) & 1.036(0.039) & -18.88(0.01) & 1.040(0.013) & -18.72(0.01) & 1.012(0.013) & 33.55& 0.168& - & (8) \\
SN1998dm & -17.30(0.12) & 1.282(0.062) & -17.16(0.03) & 1.089(0.030) & -17.34(0.03) & 1.150(0.015) & -17.54(0.02) & 1.094(0.011) & -17.55(0.01) & 1.085(0.009) & 31.86& 0.272& - & (8) \\
SN1998dx & -19.57(0.08) & 0.805(0.043) & -19.12(0.03) & 0.866(0.032) & -18.97(0.02) & 0.880(0.028) & -19.04(0.02) & 0.829(0.045) & -18.74(0.03) & 0.942(0.079) & 36.68& 0.820& - & (8) \\
SN1998ec & -18.67(0.26) & 1.176(0.109) & -18.52(0.04) & 1.029(0.023) & -18.67(0.05) & 1.008(0.024) & -18.66(0.02) & 1.086(0.013) & -18.49(0.01) & 1.048(0.018) & 34.66& 0.833& - & (8) \\
SN1998ef$^{*}$ & -19.89(0.02) & 0.987(0.038) & -19.43(0.02) & 0.904(0.012) & -19.36(0.01) & 0.953(0.008) & -19.32(0.01) & 0.976(0.016) & -19.16(0.01) & 0.881(0.005) & 34.27& 1.006& - & (8) \\
SN1998eg$^{*}$ & -19.35(0.03) & 0.954(0.058) & -18.93(0.02) & 0.970(0.031) & -18.87(0.01) & 0.974(0.031) & -18.94(0.01) & 0.918(0.050) & -18.68(0.02) & 1.927(0.340) & 35.02& 0.424& - & (8) \\
SN1998es & -19.75(0.02) & 1.146(0.023) & -19.21(0.01) & 1.132(0.016) & -19.26(0.01) & 1.112(0.008) & -19.28(0.01) & 1.102(0.015) & -18.99(0.01) & 1.137(0.012) & 33.05& 0.708& - & (8) \\
SN1999X & -19.21(0.38) & 1.075(0.133) & -19.05(0.14) & 0.938(0.072) & -18.89(0.09) & 1.007(0.074) & -18.86(0.06) & 1.090(0.034) & -18.75(0.02) & 0.928(0.064) & 35.15& 0.149& - & (8) \\
SN1999aa$^{**}$ & -19.91(0.03) & 1.205(0.025) & -19.33(0.02) & 1.143(0.026) & -19.22(0.03) & 1.156(0.054) & -19.20(0.01) & 1.136(0.014) & -18.89(0.01) & 1.122(0.006) & 34.06& 1.248& - & (8) \\
SN1999ac$^{*}$ & -19.30(0.02) & 1.092(0.026) & -19.01(0.01) & 1.015(0.010) & -18.98(0.00) & 1.069(0.004) & -19.03(0.00) & 1.044(0.004) & -18.90(0.01) & 0.991(0.003) & 33.10& 2.361& - & (8) \\
SN1999cc$^{*}$ & -19.13(0.03) & 0.994(0.036) & -18.88(0.01) & 0.850(0.012) & -18.80(0.01) & 0.887(0.012) & -18.90(0.01) & 0.843(0.010) & -18.66(0.02) & 0.812(0.011) & 35.66& 0.751& - & (8) \\
SN1999cl & -17.28(0.02) & 0.908(0.036) & -17.92(0.02) & 0.972(0.019) & -19.03(0.01) & 1.060(0.029) & -19.57(0.02) & 1.005(0.136) & -19.79(0.02) & 0.935(0.047) & 32.83& 0.631& - & (8) \\
SN1999cp$^{**}$ & - & - & -19.29(0.01) & 1.075(0.005) & -19.18(0.01) & 1.012(0.007) & -19.12(0.01) & 1.042(0.011) & -18.79(0.01) & 1.388(0.043) & 33.21& 2.162& - & (9) \\
SN1999dk$^{*}$ & - & - & -19.08(0.01) & 1.099(0.013) & -19.05(0.00) & 1.214(0.010) & -18.99(0.01) & 1.240(0.017) & -18.76(0.01) & 1.094(0.009) & 33.87& 2.868& - & (10) \\
SN1999dq & -19.94(0.01) & 1.151(0.014) & -19.43(0.00) & 1.094(0.004) & -19.45(0.00) & 1.143(0.005) & -19.49(0.01) & 1.119(0.006) & -19.28(0.01) & 1.089(0.004) & 33.82& 1.443& - & (8) \\
SN1999ee & -18.62(0.12) & 1.092(0.094) & -18.37(0.02) & 1.124(0.016) & -18.60(0.08) & 1.148(0.087) & -18.71(0.03) & 1.095(0.035) & -18.50(0.04) & 1.107(0.044) & 33.25& 1.498& - & (10) \\
SN1999ef$^{**}$ & -19.04(0.10) & 1.423(0.123) & -19.00(0.03) & 1.037(0.021) & -18.86(0.02) & 1.062(0.014) & -18.91(0.02) & 1.024(0.023) & -18.49(0.03) & 0.954(0.023) & 36.05& 1.270& - & (8) \\
SN1999ej$^{*}$ & -18.68(0.08) & 0.845(0.051) & -18.33(0.03) & 0.792(0.021) & -18.26(0.01) & 0.863(0.026) & -18.32(0.02) & 0.838(0.014) & -18.15(0.02) & 0.781(0.020) & 33.68& 0.527& E & (8) \\
SN1999ek & - & - & -18.89(0.01) & 0.954(0.010) & -18.92(0.01) & 0.990(0.012) & -18.99(0.01) & 0.935(0.014) & -18.82(0.01) & 0.892(0.006) & 34.38& 0.837& - & (10) \\
SN1999gd & -17.54(0.07) & 1.123(0.069) & -17.64(0.02) & 0.942(0.010) & -17.97(0.01) & 0.984(0.010) & -18.23(0.02) & 0.956(0.014) & -18.24(0.02) & 0.948(0.013) & 34.57& 1.518& - & (8) \\
SN1999gh & -19.01(0.12) & 0.773(0.049) & -18.62(0.02) & 0.756(0.016) & -19.12(0.03) & 0.595(0.008) & -19.07(0.01) & 0.659(0.005) & -18.82(0.01) & 0.692(0.004) & 32.86& 0.946& E & (8) \\
\hline
\end{tabular}
}
\flushleft
Table \ref{data} continued.  
\end{minipage}}
\end{figure}

\clearpage

\begin{figure}
\rotatebox{90}{% 90"x‰ñ"]'³'¹'é
\begin{minipage}{\textheight}\centering
{\small
\tabcolsep = 1mm
\begin{tabular}{lcccccccccccccl} \hline\hline
SN name & U & $s_{(U)}$ & B & $s_{(B)}$ & V & $s_{(V)}$ & R & $s_{(R)}$ & I & $s_{(I)}$ & $\mu$ & $\frac{\chi^2}{d.o.f.}$ & host & ref. \\\hline
SN1999gp$^{*}$ & -19.80(0.02) & 1.186(0.022) & -19.22(0.01) & 1.204(0.008) & -19.19(0.00) & 1.236(0.006) & -19.21(0.01) & 1.241(0.007) & -18.89(0.01) & 1.208(0.019) & 35.23& 1.849& - & (8) \\
SN2000B & -19.16(0.14) & 0.968(0.063 & -18.90(0.03) & 0.854(0.029) & -19.14(0.06) & 0.778(0.020) & -19.00(0.01) & 0.876(0.009) & -18.67(0.01) & 0.885(0.016) & 34.58& 0.638& E & (8) \\
SN2000E & -18.85(0.01) & 1.166(0.011) & -18.52(0.00) & 1.101(0.003) & -18.62(0.00) & 1.101(0.003) & -18.66(0.00) & 1.102(0.006) & -18.51(0.01) & 1.057(0.005) & 31.31& 2.699& - & (11) \\
SN2000bh$^{*}$ & - & - & -19.02(0.03) & 0.995(0.017) & -18.97(0.01) & 1.043(0.007) & -19.00(0.02) & 1.029(0.007) & -18.63(0.02) & 1.025(0.005) & 34.95& 0.383& - & (12) \\
SN2000bk & - & - & -18.22(0.01) & 0.766(0.006) & -18.44(0.01) & 0.750(0.004) & -18.50(0.00) & 0.748(0.004) & -18.35(0.01) & 0.732(0.003) & 35.18& 1.158& E & (10) \\
SN2000ca$^{**}$ & -20.08(0.03) & 1.142(0.033) & -19.46(0.01) & 1.118(0.009) & -19.33(0.01) & 1.080(0.007) & -19.29(0.01) & 1.067(0.006) & -18.90(0.01) & 1.065(0.005) & 35.02& 1.951& - & (12) \\
SN2000ce & -17.08(0.14) & 1.313(0.098) & -17.26(0.06) & 1.035(0.035) & -17.66(0.04) & 1.077(0.034) & -17.95(0.07) & 1.151(0.088) & -18.02(0.03) & 1.138(0.088) & 34.21& 0.307& - & (8) \\
SN2000cf$^{*}$ & -18.97(0.14) & 0.987(0.120) & -18.91(0.03) & 0.873(0.017) & -18.82(0.01) & 0.937(0.015) & -18.91(0.02) & 1.078(0.049) & -18.57(0.01) & 1.066(0.042) & 35.95& 0.473& - & (8) \\
SN2000cn & -18.44(0.03) & 0.974(0.025) & -18.44(0.01) & 0.761(0.009) & -18.53(0.01) & 0.835(0.011) & -18.62(0.01) & 0.808(0.019) & -18.41(0.01) & 0.769(0.017) & 34.98& 1.126& - & (8) \\
SN2000cx & -19.70(0.07) & 1.138(0.087) & -19.34(0.06) & 0.863(0.055) & -19.41(0.10) & 0.746(0.046) & -19.29(0.06) & 0.766(0.029) & -18.87(0.04) & 0.807(0.018) & 32.40& 0.220& E & (8) \\
SN2000dk$^{*}$ & -19.22(0.01) & 0.805(0.011) & -18.87(0.01) & 0.762(0.008) & -18.83(0.01) & 0.836(0.006) & -18.90(0.01) & 0.789(0.007) & -18.57(0.01) & 0.777(0.020) & 34.21& 1.792& E & (8) \\
SN2001V & - & - & -19.56(0.02) & 1.174(0.021) & -19.58(0.03) & 1.153(0.043) & -19.57(0.02) & 1.151(0.024) & -19.35(0.02) & 1.123(0.025) & 34.17& 1.319& - & (13) \\
SN2001ba$^{**}$ & - & - & -19.41(0.01) & 1.043(0.008) & -19.28(0.01) & 1.030(0.012) & - & - & -18.85(0.01) & 1.016(0.010) & 35.61& 0.888& - & (12) \\
SN2001bt & - & - & -18.71(0.01) & 0.899(0.007) & -18.82(0.00) & 1.006(0.003) & -18.95(0.01) & 0.962(0.009) & -18.75(0.01) & 0.954(0.009) & 33.97& 0.630& - & (3) \\
SN2001cn & - & - & -18.88(0.01) & 0.950(0.007) & -18.94(0.01) & 1.026(0.003) & -19.04(0.01) & 0.969(0.004) & -18.77(0.01) & 0.975(0.004) & 34.10& 0.561& - & (3) \\
SN2001cz & - & - & -19.18(0.01) & 1.056(0.012) & -19.22(0.01) & 1.058(0.008) & -19.26(0.01) & 0.998(0.017) & -18.98(0.01) & 1.039(0.010) & 34.23& 0.810& - & (3) \\
SN2001el & -18.20(0.01) & 0.993(0.017) & -18.14(0.01) & 0.953(0.012) & -18.20(0.01) & 1.090(0.014) & -18.38(0.01) & 0.993(0.006) & -18.09(0.01) & 1.060(0.011) & 30.92& 1.128& - & (14) \\
SN2001en & -18.83(0.13) & 1.241(0.436) & -18.79(0.01) & 0.953(0.007) & -18.86(0.01) & 1.027(0.007) & -18.96(0.01) & 0.971(0.004) & -18.69(0.01) & 0.976(0.005) & 34.03& 0.542& - & (15) \\
SN2002bo & - & - & -17.92(0.02) & 0.951(0.024) & -18.25(0.01) & 0.991(0.009) & -18.51(0.04) & 1.007(0.027) & -18.37(0.01) & 0.997(0.007) & 31.85& 0.907& - & (3) \\
SN2004S & - & - & -18.93(0.02) & 0.946(0.016) & -18.78(0.03) & 1.103(0.017) & -18.88(0.02) & 0.990(0.012) & -18.65(0.02) & 0.997(0.011) & 33.00& 0.479& - & (16) \\\hline
\end{tabular}
}
\flushleft
Table \ref{data} continued.  
\end{minipage}}
\end{figure}

\section{RESULTS}

Figure \ref{day_chi} shows the size of typical residuals from the template light curves in U, B, V, R and I-band. Most of observed light curves follow the templates closely. However, note that some early phase U-band photometry (day $<$ 0) is systematically brighter than the template. This could be due to K-corrections (we made the templates by combining observed light curves without K-corrections, see \S3.4).

However, there were a few SNe Ia which deviated significantly from the templates, due to an intrinsic peculiarity in their light curve shapes or a failure to estimate their photometric errors properly. Peculiar light curves which don't follow the templates are very interesting for studies of the nature of supernovae, but that is not our purpose in this paper. Therefore, we exclude 14 / 122 SNe Ia with reduced $\chi^2$ values larger than 3 (see also Figure \ref{chi_count}) from further analysis. We did not exclude SNe Ia due to any other properties such as spectroscopic subtypes or host galaxy types. Only the exception is that we discuss those peculiar SNe in \S4.3 related to I-band light curves. We refer to the remaining 108 / 122 SNe Ia in the following discussions All parameters of this subset of 108 SNe Ia are given at Table \ref{data}. As we wrote in \S2, all of 108 SNe have B and V-band photometry, but they don't always have U, R and I-band photometry (see Table \ref{sample}).

We also make some subsets of SNe Ia based on colours, stretch factors, host type and redshift in the following discussion. Those subsets are all made from the selected 108 SNe Ia.

\subsection{Stretch Factor}

In Figure \ref{sf_count}, we show distributions of U, B, V, R and I-band stretch factors. The values in V, R and I-band are more concentrated than those in U and B-band. Since the U-band stretch factor suffers from larger errors than the B-band factor, due to much fewer measurements and larger photometric errors, we conclude that the B-band stretch factor is the best choice to serve as an index of SN Ia properties; we refer to it extensively in the following sections. Actually, we use the reciprocal of the B-band stretch factor in this paper in order more easily to compare our results with those of previous authors who used $\Delta m_{15}$ as a light curve shape parameter.

In Figure \ref{Bsf_asf_dm15}, we compare our stretch factors with those of ALT04, using the common set of 58 SNe Ia. We obtain  $s_{(B)Altavilla}^{-1} = (0.86 \pm 0.07) \times s_{(B)}^{-1} + (0.13 \pm 0.07)$. The disagreement between our stretch factors and those of ALT04 may be due to the difference in time coverage of templates used in the light curve fitting. We find the relationship between $\Delta m_{15}(B)$ and the inverse of our B-band stretch factor to be $\Delta m_{15} = (1.67 \pm 0.12)  \times (s_{(B)}^{-1} - 1) + (1.13 \pm 0.02)$. This result is consistent with ALT04, which found $\Delta m_{15} = (1.98 \pm 0.16)  \times (s_{(B)Altavilla}^{-1} - 1) + (1.13 \pm 0.02)$.

Figure \ref{Bsf_sf} shows the relationship between the inverse B-band stretch factor and the inverse U, V, R and I-band stretch factors. The best fit relations are listed in Table \ref{Bsf_sf_tab}. JHA06 showed that relations among U, B and V-band stretch factors are consistent with a "universal" stretch, and our results agree: our B, V, R and I-band stretch factors are equal within the formal uncertainties. However, the coefficient of the relationship between U and B-band stretch factors is smaller than those for the relationships between B, V, R and I-band stretch factors. If we select 19 SNe Ia in common with JHA06, the relation between U and B-band stretch factors is $s_{(U)} = (1.03 \pm 0.11) \times s_{(B)} + (0.05 \pm 0.11)$. This result is consistent with those of JHA06 ($s_{(U)} = 1.04 \times s_{(B)} - 0.02$).

\begin{table*}
\begin{center}
\begin{minipage}{30em}
\caption{Relationships between inverse B-band stretch factor and stretch factors in U,V, R and I-band}
\label{Bsf_sf_tab}
\end{minipage}
\small
\begin{tabular}{cccc} \hline\hline
Band & Relation & r.m.s. (mag) & Number \\ \hline
U-band & $(0.78 \pm 0.07) \times s_{(B)}^{-1} + (0.14 \pm 0.07)$ & 0.040 & \phantom{0}44 \\
V-band & $(0.93 \pm 0.05) \times s_{(B)}^{-1} + (0.03 \pm 0.05)$ & 0.009 & 108 \\
R-band & $(1.04 \pm 0.05) \times s_{(B)}^{-1} - (0.05 \pm 0.05)$ & 0.009 & \phantom{0}88 \\
I-band & $(1.02 \pm 0.05) \times s_{(B)}^{-1} - (0.01 \pm 0.06)$ & 0.011 & 102 \\ \hline
\end{tabular}
\end{center}
\end{table*} %

\subsection{Magnitude and colour}

\begin{table}
\begin{center}
\begin{minipage}{30em}
\caption{Average absolute magnitudes (r.m.s.) and stretch factors (r.m.s.)}
\label{mag_sf}
\end{minipage}
\small
\begin{tabular}{ccccc} \hline\hline
 & Magnitude & Stretch Factor & Number \\ \hline
U-band & \llap{$-$}19.08 (0.79) & 1.07 (0.14) & \phantom{0}44 \\
B-band & \llap{$-$}18.79 (0.76) & 0.96 (0.12) & 108 \\
V-band & \llap{$-$}18.84 (0.58) & 1.00 (0.13) & 108 \\
R-band & \llap{$-$}18.93 (0.53) & 1.00 (0.13) & \phantom{0}88 \\
I-band & \llap{$-$}18.70 (0.45) & 0.98 (0.15) & 102 \\ \hline
\end{tabular}
\end{center}
\end{table} %

Figure \ref{mag_count} shows the distribution of U, B, V, R and I-band absolute peak magnitudes corrected for Galactic extinction according to \cite{sch98}. The trend that redder bands are more concentrated than bluer bands possibly indicates the effects of extinction by dust (see also Table \ref{mag_sf}).

The colours of the SNe Ia also suggest that dust plays an important role. Figure \ref{BV_count} shows the distribution of the $(B-V)$ colour at the time of B-band maximum, $(B-V)_{max}$. The tail of redder side in Figure \ref{BV_count} are likely SNe Ia reddened by dust in their host galaxies. As shown in Figure \ref{z_BV}, redder SNe Ia tend to be missing at higher redshift, implying that the redder SNe Ia are affected by dust. The clear edges of the blue side of the distributions in Figures \ref{BV_count}, and \ref{z_BV} indicates that the $(B-V)_{max}$ colour of bluest SNe Ia is almost constant. Fitting a single-sided gaussian to the left side of the histogram in Figure \ref{BV_count} ($(B-V)_{max} > -0.20$), we estimate the intrinsic colour of bluest SN Ia as $(B-V)_{max} = -0.12$. Figure \ref{BV_colour} shows relationships between $(B-V)_{max}$ and the other colours at the time of B-band maximum. We will discuss this in detail in \S5. 

We should pay attention to the possibility that there are intrinsically redder SNe Ia. We cannot distinguish those redder SNe Ia from the SNe Ia reddened by host galaxy dust based on apparent colour. In \S6 and \S7, we discuss about the issue.

Based on the the result of single-side gaussian fitting of B-V distribution, we made three subsamples, ``BV bluest'' SNe Ia ($-0.14 < (B-V)_{max} \le -0.10$, $-0.12 \pm 0.02$ mag around the median), ``BV bluer'' SNe Ia ($-0.10 < (B-V)_{max} < -0.02$, 0.1 mag redder than the median) as less dust affected samples and ``BV redder'' SNe Ia ($0.00 < (B-V)_{max}$) as a dust affected sample. We use them in the following discussions.

\subsection{I-band Peculiarity}

It is well known that there is quite a variety of shapes in the I-band light curves of SNe Ia. Usually, there is a ``bump'' or a ``second peak'' in the I-band, but some SNe Ia don't have a secondary peak (see Figure \ref{IbandLC}). Some of previous authors fitted these I-band light curves using two Gaussian functions (\citealt{con00,nob05}). However, we tried to fit I-band light curves using only one template to avoid introducing extra parameters.

From all of 122 SNe Ia, we selected 71 well observed SNe Ia which have at least one I-band measurement in two periods, -10 days to 0 days and +10 days to +20 days. Based on photometry during these periods, the 71 SNe Ia were classified into two types, normal I-band light curves and peculiar I-band light curves (see Figure \ref{Ibandclass}), where ``peculiar'' means that the I-band light curves are fainter than the I-band template from -10 days to 0 days and brighter than the I-band template from +10 days to +20 days.

We found about 90\% of SNe Ia (63/71 SNe) can be fitted well with the template and only 8 SNe Ia (SN1991bg, SN1995ac, SN1997br, SN1998ab, SN1999ac, SN1999by, SN1999gh, SN2002cx) can not be fitted well with the template. Most of those which deviate from the template are known to be spectroscopically peculiar as well. For example, SN1991bg and SN1999by are classified as ``SN1991bg-like,'' SN1995ac and SN1997br are classified as ``SN1991T-like,'' and SN2002cx is known as very peculiar SNe Ia \citep{li03}. On the other hand, I-band light curves of SN1991T and SN1995bd, which are classified spectroscopically as ``SN1991T-like'' do fit the I-band light template. Based on the value of reduced $\chi^2$ calculated with Multi-band Stretch method, these SNe which do not fit the template I-band light curve have been excluded from the samples in the following discussions, as we mentioned in the beginning of \S4. As \cite{nob05} pointed out the correlation between lack of a second I-band maximum and a lower B-band stretch, the Multi-band Stretch Method would have room for improvement in the treatment of these SNe.

\section{SUB GROUP OF TYPE Ia SUPERNOVAE}

Do all SNe Ia with the same B-band stretch factor have the same intrinsic colours? The best way to answer this question is to compare a series of well-calibrated spectra from different events, but we lack such a dataset. Therefore, we have investigated the issue using photometric information from our sample of SNe Ia.

In Figure \ref{Bsf_VI}, we plot SNe Ia classified on the basis of their $(B-V)_{max}$ colour as ``BV bluest'' SNe Ia, ``BV redder'' SNe Ia and ``others''. As we discussed in \S4, the bluer SNe Ia may be free from dust extinction. In Figure \ref{Bsf_VI}, there is a "sub" group at the lower right, with inverse B-band stretch factor$ > 1.1$ and $(V-I)_{max} < 0$. This "sub" group consists of SNe Ia which are redder in $(B-V)_{max}$, though they are bluer in $(V-I)_{max}$. It is difficult to explain this combination of colours using ordinary SNe Ia shining through dust in their host galaxies. Instead, it may mean that the ``sub'' group is due to a set of SNe Ia with intrinsic SEDs which differ from the ``main'' group with inverse B-band stretch factor$ < 1.1$, at least in B, V and I-band. 

There are some observational indications that properties of SNe Ia are related to their environment. \cite{ham96} showed that the mean luminosity of SNe Ia in elliptical galaxies is fainter than that of SNe Ia in spiral galaxies. \cite{man05} and \cite{sul06} also showed that the star-forming activity in a galaxy is related to properties of its SNe Ia. We have plotted the SNe Ia hosted by E or S0 galaxies in Figure \ref{Bsf_colour}. We note that the SNe Ia of the ``sub'' group at the lower right tend to be found in E or S0 galaxies. However, not all SNe Ia found in E or S0 galaxies belong to this group.

Of course, a difference of intrinsic colours doesn't mean a difference of intrinsic luminosity directly. However, a difference in intrinsic colours affects the accuracy of correcting for host galaxy dust extinction if we correct the extinction according to apparent colour of SN Ia. For example, the bluer SNe Ia with inverse B-band stretch factors $ > 1.2$ in the $(B-V)$ panel of Figure \ref{Bsf_colour} are redder than the bluer SN Ia with inverse B-band stretch factors $ < 1.2$ by about 0.05 magnitude in $(B-V)_{max}$. If we treat the $(B-V)_{max}$ as a constant when we estimate the host galaxy dust extinction, the correction will be overestimated systematically by about 0.2 magnitude in B-band. This is critical if we want to use SNe Ia as distance indicators for cosmological studies. 

\section{RELATIONS BETWEEN B-BAND STRETCH FACTOR AND BRIGHTNESS}

Figure \ref{Bsf_Bmax} and Figure \ref{Bsf_UVRImax} show the relationships between inverse B-band Stretch factor and luminosity at the B-band maximum (hereafter, stretch - magnitude relation) in U, B, V, R and I-band. The scatter in the luminosity is mainly caused by three factors: the peculiar velocity of the host galaxy, extinction by dust in the host galaxy, and intrinsic diversity of SNe Ia.

Peculiar velocity makes the dispersion larger when we estimate the distance to a SN from its host's recession velocity. The effect of peculiar velocity must be treated properly if the host galaxy is nearby. For example, if a host galaxy has peculiar velocity of 400 km/s, we will overestimate or underestimate the luminosity of its SN by $\sim$0.3 mag at $z = 0.01$ and $\sim$0.1 mag at $z = 0.02$. However, statistically speaking, the effect can be small if the sample size is large enough. In our case, Table \ref{Bmag_redshift} shows that there is no significant effect on averaged absolute magnitudes caused by peculiar velocity if we choose ``BV bluest'' + ``BV bluer'' SNe Ia in order to avoid the effect of host galaxy dust extinction. In Figure \ref{z_max}, we plotted individual SNe Ia shown in Table \ref{Bmag_redshift}, however there are no considerable effect caused by peculiar velocities. Table \ref{Bmag_redshift} also shows the scatter size in the nearest sample ($0.00 < z \le 0.02$) is only 10\% larger than the further samples.

It is better to include properly the effect of peculiar velocity in error bars, however we don't know which value could represent peculiar velocities. For example, peculiar velocity of galaxies in the clusters often reaches $\sim$1,000km/s. It is very difficult to estimate appropriate value of peculiar velocities without additional observations, and we don't want to unnecessarily overestimate the size of error bars.

One way to avoid the peculiar velocity problem is to use further redshift SNe Ia only. But, the number of SNe Ia becomes smaller in compensation for the selection. Hence, we didn't make any corrections for the effect of peculiar velocity in this study.

\begin{table}
\begin{center}
\begin{minipage}{30em}
\caption{Average magnitudes of ``BV bluest'' + ``BV bluer'' SNe Ia at different redshifts}
\label{Bmag_redshift}
\end{minipage}
\small
\begin{tabular}{ccc} \hline\hline
Redshift & $M_{B}$ & Number \\ \hline
$0.00 < z \le 0.02$ & \phantom{0}\llap{$-$}19.13 (0.35) & 18 \\
$0.02 < z \le 0.05$ & \phantom{0}\llap{$-$}19.15 (0.23) & 19 \\
$0.05 < z \phantom{000000}$ & \phantom{0}\llap{$-$}19.01 (0.26) & 10 \\ \hline
\end{tabular}\\
\vspace*{0.6em}
{\footnotesize
\begin{minipage}{30em}
\end{minipage}}
\end{center}
\end{table} %

Extinction due to dust in the host galaxy is a major contributor to the scatter in luminosity. Unfortunately, it is very difficult to estimate this extinction properly. We must separate the effect of host galaxy dust extinction from intrinsic colour variations of SNe Ia. One way to avoid this problem is to use only bluer SNe Ia which are expected to be free from extinction. In Figure \ref{Bsf_Bmax}, we plot the ``BV bluest'' SNe Ia and ``BV bluer'' SNe Ia samples. Coefficients of the relationship between absolute magnitude and inverse stretch factor are listed in Table \ref{Bsf_mag_coe}. In all passbands, the slope of the line derived from the ``BV bluest'' sample is about double that of the slope of the line derived from the combined ``BV bluest'' + ``BV bluer'' sample. Figure \ref{Bsf_Bmax} and Figure \ref{Bsf_UVRImax} also show that most of ``BV bluest'' SNe have broader light curve widths than ``BV bluer'' SNe. This suggests there may be different groups of SNe Ia with different intrinsic properties. The slope of the stretch - magnitude relation of ALT04 is 1.83 (using 1.102 in Table 1 of Altavilla's paper and $\Delta m_{15} = (1.67 \pm 0.12)  \times (s_{(B)}^{-1} - 1) + (1.13 \pm 0.02)$), which is shallower than the slope of ``BV bluest'' SNe Ia but steeper than the slope of the combined ``BV bluest'' + ``BV bluer'' sample.

We also tried to compare the dispersion of luminosity of ``BV bluest'' SNe Ia around the stretch - magnitude relation derived from ``BV bluest'' SNe Ia with that of SNe Ia hosted by E or S0 galaxies. Those SNe Ia hosted by E or S0 galaxies should be largely free from extinction due to interstellar dust in the host galaxy, but Figure \ref{Bsf_Bmag_ES0} shows that they have a larger dispersion than ``BV bluest'' SNe Ia (which are also expected to be free from host galaxy dust extinction); see also Table \ref{Bsf_Bmag_type}. Even if we select ``BV bluer'' SNe Ia hosted by E or S0 galaxies, the dispersion is still larger than ``BV bluest'' SNe Ia without any dust / colour corrections. These facts may indicate that SNe Ia hosted by E or S0 galaxies have larger intrinsic diversity than SNe Ia hosted by other type galaxies, or that the host galaxies have some significant dust extinction.

Of course, the scatter of brightness of SNe Ia in E or S0 galaxies could be attributed to the peculiar velocity of the host galaxies, but the peculiar velocity of host galaxies can not account for the fact that direction of distribution of SNe Ia hosted by E or S0 galaxies shown in Figure \ref{BV_colour} is parallel to the direction of distribution due to dust in the Milky Way.

\subsection{Colour Correction For Absolute Magnitude}

It is known empirically that correcting the luminosity of SNe Ia based on their colours will decrease the dispersion of peak B-band luminosities. For example, MLCS, CMAGIC, and SALT use colour information as part of the light curve fitting technique, while Stretch and $\Delta m_{15}$ methods use colour information to decrease the dispersion after fitting light curves. We also introduce colour terms to decrease the dispersion around the stretch - magnitude relation.

We need a colour excess to remove the extinction due to dust in the host galaxy, and the intrinsic colour of each SN Ia to correct for diversity among SNe Ia. However, it is difficult to separate an observed colour excess into the intrinsic colour of the SN and the reddening by dust. Therefore, we don't try to distinguish them; instead, we use the observed colour of SNe Ia empirically to reduce the dispersion of the relationship between absolute magnitude and inverse stretch factor. We employ a chi-square fitting technique with following equation.

\begin{equation}
\chi = M_{B} - \{ a \times s_{(B)}^{-1} + b \times (B - V)_{max} + c \}
\end{equation}

Here $M_{B}$ is the absolute B-band magnitude at maximum shown in Table \ref{data} without any corrections, $(B-V)_{max}$ is the apparent colour at B-band maximum and $a, b, c$ are fitting parameters. Note that here we treat $(B-V)_{max}$ as independent from B-band stretch factor.

Table \ref{Bsf_Bmag_colour} shows the inverse B-band stretch factor and B-band luminosity relationships, including the r.m.s. of each sample. The r.m.s. is the smallest in sample 8D, which includes SNe Ia hosted by E or S0 galaxies at higher redshift (z $>$ 0.02). The relationship of sample 8D is similar to that of sample 8B (SNe at $z > 0.02$), but the r.m.s. is smaller in the sample drawn from E or S0 galaxies. This may indicate that extinction by dust in an E or S0 host galaxy is smaller than extinction in other types of galaxies, if both sample 8B and 8D have the same intrinsic dispersion of peak B-band luminosity. 

\begin{table*}
\begin{center}
\begin{minipage}{40em}
\caption{Relationships between B-band stretch factor and luminosity}
\label{Bsf_mag_coe}
\end{minipage}
\small
\begin{tabular}{ccccc} \hline\hline
 & Sample & Relation & r.m.s. (mag) & Number \\ \hline
U-band & 6A & $(3.28 \pm 0.40) \times s_{(B)}^{-1} - (22.93 \pm 0.42)$ & 0.32 & \phantom{0}8 \\
 & 6B & $(1.50 \pm 0.40) \times s_{(B)}^{-1}.- (21.11 \pm 0.44)$ & 0.32 & 18 \\
B-band & 6A & $(2.28 \pm 0.42) \times s_{(B)}^{-1} - (21.49 \pm 0.41)$ & 0.17 & 21 \\
 & 6B & $(1.08 \pm 0.24) \times s_{(B)}^{-1} - (20.30 \pm 0.24)$ & 0.23 & 46 \\
V-band & 6A & $(2.58 \pm 0.38) \times s_{(B)}^{-1} - (21.67 \pm 0.38)$ & 0.20 & 21 \\
 & 6B & $(1.10 \pm 0.24) \times s_{(B)}^{-1} - (20.22 \pm 0.25)$ & 0.22 & 46 \\
R-band & 6A & $(1.88 \pm 0.57) \times s_{(B)}^{-1} - (20.96 \pm 0.56)$ & 0.18 & 14 \\
 & 6B & $(0.75 \pm 0.29) \times s_{(B)}^{-1} - (19.89 \pm 0.30)$ & 0.26 & 34 \\
I-band & 6A & $(1.10 \pm 0.61) \times s_{(B)}^{-1} - (19.93 \pm 0.60)$ & 0.23 & 20 \\
 & 6B & $(0.52 \pm 0.30) \times s_{(B)}^{-1} - (19.41 \pm 0.31)$ & 0.27 & 42 \\ \hline
\end{tabular}\\
\vspace*{0.6em}
{\footnotesize
\begin{minipage}{40em}
6A is "BV bluest" sample.\\
6B is "BV bluest"+"BV bluer" sample.
\end{minipage}}
\end{center}
\end{table*} %

\begin{table}
\begin{center}
\begin{minipage}{30em}
\caption{r.m.s. of the stretch - magnitude relation of the samples and their average $(B-V)_{max}$}
\label{Bsf_Bmag_type}
\end{minipage}
\small
\begin{tabular}{cccc} \hline\hline
Sample & r.m.s. (mag) & average $(B-V)_{max}$ & Number \\ \hline
7A & 0.17 & -0.12 & 21 \\
7B & 0.51 & \phantom{-}0.04 & 27 \\
7C & 0.30 & -0.07 & 15 \\
7D & 0.17 & -0.12 & 3 \\ \hline
\end{tabular}\\
\vspace*{0.6em}
{\footnotesize
\begin{minipage}{30em}
7A is "BV bluest" sample.\\
7B is sample hosted by E or S0 galaxies with no colour selection.\\
7C is "BV bluer" sample hosted by E or S0 galaxies.\\
7D is "BV bluest" sample hosted by E or S0 galaxies.
\end{minipage}}
\end{center}
\end{table} %

\begin{table*}
\begin{center}
\begin{minipage}{40em}
\caption{Relationships between B-band inverse stretch factor and luminosity with colour term and r.m.s. of the stretch - magnitude relation}
\label{Bsf_Bmag_colour}
\end{minipage}
\small
\begin{tabular}{cccc} \hline\hline
Sample & relation & r.m.s. (mag) & Number \\ \hline
8A & $(0.96 \pm 0.28) \times s_{(B)}^{-1}- (2.51 \pm 0.15) \times (B - V)_{max} - (20.26 \pm 0.29)$ & 0.48 & 108 \\
8B & $(0.98 \pm 0.18) \times s_{(B)}^{-1}- (2.28 \pm 0.28) \times (B - V)_{max} - (19.95 \pm 0.19)$ & 0.27 & \phantom{0}50\\
8C & $(1.09 \pm 0.59) \times s_{(B)}^{-1}- (1.78 \pm 0.43) \times (B - V)_{max} - (20.15 \pm 0.71)$ & 0.33 & \phantom{0}27 \\
8D & $(0.99 \pm 0.34) \times s_{(B)}^{-1}- (2.23 \pm 0.24) \times (B - V)_{max} - (20.10 \pm 0.41)$ & 0.12 & \phantom{0}15 \\
\hline
\end{tabular}\\
\vspace*{0.6em}
{\footnotesize
\begin{minipage}{40em}
8A is all SNe Ia.\\
8B is SNe Ia of z $>$ 0.02.\\
8C is SNe Ia hosted by E or S0 galaxies.\\
8D is SNe Ia of z $>$ 0.02 hosted by E or S0 galaxies.\\
\end{minipage}}
\end{center}
\end{table*} %

\section{HOST GALAXY DUST PROPERTIES}

There are several approaches to estimate the effects of extinction by dust in host galaxies. In this section, we estimate host galaxy dust extinction using the apparent colours of SNe Ia. Dust in a host galaxy dust makes the apparent colour of a SN Ia redder, so we can estimate how a host galaxy obscures a SN using colour information. We use the following relationship:

\begin{equation}
A_{X} = R_{X} \times E(M_{B}-M_{V})
\end{equation}

Here $A_{X}$ is dust absorption in X-band, $R_{X}$ is the ratio of total to selective extinction in X-band, and $E(M_{B}-M_{V})$ is the colour excess. We derive $E(M_{B}-M_{V})$ assuming an intrinsic $(B-V)_{max}$ colour for each SN Ia and extract the excess from observed colour.

There are two main problems when we use equation (4) to correct for extinction in the host galaxy. One problem is that we don't know if the properties of dust in host galaxies are the same as those of dust in the Milky Way, especially the value of $R_{X}$. ALT04 and other works showed that $R_{B}$ of dust in host galaxies may be smaller than $R_{B}$ of dust in the Milky Way. Another problem is that we don't know the intrinsic colour of SNe Ia. Usually we assume that the bluest SNe Ia are free from extinction, but there is no guarantee that all SNe Ia have the same intrinsic colour as the bluest examples. These two problems make the correction difficult. Hence, we reduce the problem to two extreme cases in the following.

In case A, we assume that the intrinsic colour of SNe Ia depends on the inverse B-band stretch factor. This assumption was also used in ALT04. We found earlier that the intrinsic $(B-V)_{max} = -0.30 \times s_{(B)}^{-1} + 0.18$ from the relations between $s_{(B)}^{-1}$ and B, V-band (see Table \ref{Bsf_mag_coe}). This assumption is consistent with SNe Ia with $s_{(B)}^{-1} < 1.1$, but not consistent with SNe Ia with $s_{(B)}^{-1} > 1.1$ (see Figure \ref{Bsf_colour}).

In case B, we assume the intrinsic colour of SNe Ia is constant, and has some significant scatter. In this case, even if two SNe Ia have the same inverse B-band stretch factor, they don't necessarily have the same colour.

First, we tried to correct for extinction due to dust in host galaxies under the assumption of case A, using three different values for $R_{B}$: 4.315 (appropriate for the Milky Way, \citealt{sch98}), 3.5 (ALT04) and 3.0. We selected redder SNe Ia, which are $(B-V)_{max} > -0.02$ (0.1 mag redder from average ``intrinsic'' $(B-V)_{max} = -0.12$ derived from Figure \ref{BV_count}). These SNe Ia must be significantly reddened by dust in the host galaxy. Table \ref{dust_correction_01} shows the scatter around the stretch - magnitude relationship in B-band luminosity after we have corrected for extinction according to equation (4). Though smaller values of $R_{B}$ do decrease the dispersion, the scatter is still large. This means we must introduce another factor to account for the large dispersion; for example, some intrinsic scatter of $(B-V)_{max}$. Note that the r.m.s. of all samples after our corrections for dust in the host galaxy is larger than the r.m.s. shown in ALT04, Figure 5 (b) ($R_B = 3.5, \sigma = 0.31$). The explanation is mainly that we used only redder SNe in our analysis, but the ALT04 sample includes bluer SNe Ia which have smaller scatter in B-band luminosity than redder SN Ia. If we use a set of 55 SNe Ia in common with ALT04 to derive the dispersion, we find $\sigma = 0.44$ mag with $R_{B} = 0.35$, which is smaller than that of redder SN Ia ($R_B = 3.5, \sigma = 0.81$, Table \ref{dust_correction_01}) but still larger than the result shown in ALT04, Figure 5 (b) ($R_B = 3.5, \sigma = 0.31$). The reason may be in the details of how one derives E(B-V): ALT04 used a weighted average of $(B-V)_{max}$ and $(B-V)_{tail}$, but we used $(B-V)_{max}$ only.

\begin{table}
\caption{Dispersion of B-band luminosities (mag) of selected redder SNe Ia after host galaxy dust correction based on three different conversion factors of $R_{B}$.} \label{dust_correction_01}
\begin{center}
\small
\begin{tabular}{ccccc} \hline\hline
Sample & $R_{B} = 4.315$ & $R_{B} = 3.5$ & $R_{B} = 3.0$ & Number \\ \hline
all & 0.89 & 0.81 & 0.79 & 53 \\
9A & 0.99 & 0.90 & 0.89 & 38 \\
9B & 0.56 & 0.48 & 0.45 & 15 \\
ALT04 & 0.52 & 0.44 & 0.42 & 55 \\ \hline
\end{tabular}\\
\vspace*{0.6em}
{\footnotesize
\begin{minipage}{28em}
9A is "broader" LC sample, which $s_{(B)}^{-1} < 1.1$.\\
9B is "narrower" LC sample, which $s_{(B)}^{-1} > 1.1$.\\
ALT04 is common 55 samples with ALT04, Figure 5 (b).
\end{minipage}}
\end{center}
\end{table} %

Next, we tried to correct host galaxy dust extinction under the assumption of case B (a constant intrinsic colour for all SNe Ia) using a factor $R_{B} = 4.315.$ Indeed, according to Figure \ref{BV_colour}, there is no basis to assume that dust in host galaxies has properties different from dust in the Milky Way. Under the assumption of case B, we can't derive $A_{X}$ from equation (4) because we don't know the intrinsic $(B-V)_{max}$ of SNe Ia. Instead, we tried to derive $E(M_{B}-M_{V})$ from the following equation (5), under the assumption that the intrinsic B-band luminosity of SNe Ia is given by the relationships listed in Table \ref{Bsf_mag_coe}.

\begin{equation}
E(M_{B}-M_{V}) = \frac{M_{B} - (2.28 \times s_{(B)}^{-1} - 21.49)}{R_{MW}}
\end{equation}

Here we regard residuals from the intrinsic B-band luminosity at some particular B-band stretch factor as due completely to extinction by dust in the host galaxy, and we divide the residuals by $R_{MW} = 4.315$, the ratio of total to selective extinction for dust in the Milky Way. We applied equation (5) to dust-affected SNe Ia, corrected them for reddening, then compared their $(B-V)_{max}$ distribution with that of dust-free SNe Ia.

To make the dust-affected sample, we selected SNe Ia which are fainter by at least 1.0 magnitude in B-band than the intrinsic luminosity derived from inverse B-band stretch factor. For the dust-free sample, we selected SNe Ia which fall within 0.17 magnitude (which is one standard deviation of ``BV bluest'' SNe Ia) of the relationship between intrinsic luminosity and inverse B-band stretch factor. 

The results are distributions of ${(B-V)}_{max}$ colours for dust-free samples, dust-affected samples, and dust-corrected samples, using B-band residuals to correct for extinction (Figure \ref{BV_count_correct_B}), or V-band residuals to correct for extinction (Figure \ref{BV_count_correct_V}). We compare the distribution of SNe Ia which are corrected using the factor $R_{B}$ of the Milky Way to the distribution of SNe Ia which are corrected with a smaller factor $R_{B}$. These comparisons show that if we accept the idea that intrinsic SNe Ia colours have an intrinsic dispersion which is similar to that of the ``reference'' samples in Figure \ref{BV_count_correct_B} and \ref{BV_count_correct_V}, the average factor $R$ of dust in host galaxies is not different than that for dust in the Milky Way. We ran a Kolmogorov-Smirnov test on the distribution of corrected $(B-V)$ colours and found no significant difference between them. Intrinsic diversity among SNe Ia diversity has also been shown by previous spectroscopic studies. \cite{ber06}, for example, showed that r.m.s. of B-band and V-band luminosities is about 0.1 - 0.2 magnitude, and their results are consistent with ours.

If the intrinsic colours of SNe Ia have a significantly large scatter, it is very difficult to correct properly for extinction by dust in host galaxies 
with photometric measurements only. 

\section{SN Ia SAMPLES FOR COSMOLOGY}

In \S6, we studied the connection between light curve shape and luminosity and looked for ways to decrease the dispersion around the relationship. In \S7, we checked the properties of dust in host galaxies.

These results suggest that if one wishes to use SN Ia as standard candles, one should take one of the following approaches. First, one might select only ``BV bluest'' SNe Ia. As we showed in \S6.1, colour corrections make the dispersion smaller, but the improvement in the dispersion of a combined sample of ``BV bluer'' and ``BV bluest'' SNe Ia after colour corrections is only 0.02 mag, and their r.m.s. is still larger than that of the ``BV bluest'' sample with no colour correction. In addition, Figure \ref{Bsf_colour} shows that the intrinsic colour of SNe Ia seems to have a large scatter, and that even SNe which have same B-band stretch factor may not have same intrinsic colour. Until we solve two puzzles, (i) how to separate the apparent colour of SNe Ia into intrinsic colour and reddening due to the host galaxy, (ii) the value of the ratio of total to selective extinction $R$ and its evolution from the low-z to the high-z universe, we had better not use colour corrections in samples which have large colour diversity. The study of K-corrections is very important for selecting ``BV bluest'' SN Ia and determining extinction by dust in host galaxies.

Another approach is to select SNe Ia which occur in E or S0 galaxies. As shown in \S6.1, the r.m.s. of stretch - magnitude relationship of colour-corrected SNe Ia hosted by distant E or S0 galaxies (8D, Table \ref{Bsf_Bmag_colour}) is 0.12 mag, which is smaller than that of the ``BV bluest'' SNe Ia discussed above. However, the properties of SNe Ia in E or S0 galaxies tend to differ from those of SNe Ia within other types of galaxies (cf. Figure \ref{Bsf_VI}, Figure \ref{Bsf_Bmag_ES0}, and Figure \ref{Bsf_colour}). This means we may have to apply different relations to SNe Ia hosted by E or S0 galaxies. 

Even if we select SNe Ia within early type galaxies, those SNe are not always free from dust, as shown in Figure \ref{BV_colour}. This is consistent with the idea that not only interstellar dust but also dust surrounding the SN affects the light from SNe Ia, as \cite{wan05} pointed out. \cite{pat07} reported the spectroscopic detection of circumstellar dust around normal SN Ia. The SNe Ia occuring in early type galaxies also tend to have narrower light curve widths, with inverse B-band stretch factor $> 1.1$ (see Figure \ref{Bsf_colour}), and they have larger intrinsic dispersion in SED as shown in Figure \ref{Bsf_colour}. Figure \ref{Bsf_Bmag_ES0} shows these SNe Ia in early-type galaxies have larger dispersion than ``BV bluest'' SNe Ia, or that the stretch - magnitude relation is different from that of ``BV bluest'' SNe Ia. Our results are consistent with the distribution of B-band stretch factors of nearby SNe Ia within E or S0 hosts determined in \cite{sul06}. However, \cite{sul03} also claimed that at high redshifts, the scatter in luminosity of SNe Ia hosted by E or S0 galaxies is smaller than that of SNe Ia inside spiral galaxies.

In any case, we must know whether or not the properties of SNe Ia in the low-z universe and the high-z universe are the same, since it is important when we apply stretch-brightness relationships or make K-corrections. To answer these questions, we need to study the nature of SNe Ia at all redshifts.

\section{SUMMARY}

In this work, we re-analyze U, B, V, R and I-band light curves of 122 previously-published nearby (redshift $<$ 0.11) SNe Ia, which are mainly from JHA06, \cite{rie99a} and \cite{ham96}. We devise the ``Multi-band Stretch method'', which is a revised Stretch method extended to fitting up to five bands. We create new U, B, V, R and I-band light curve templates from well-observed SNe Ia for our ``Multi-band Stretch method''. It is known that the I-band light curves of SNe Ia shows much variation. However, we found our I-band template fitted about 90\% of SNe Ia I-band light curves very well.

Using the results of our light curve fitting, we examine the relationships between luminosity, colours at maximum light, and stretch factors. In \S5, we show a possible sub-group of SNe Ia which have different colours from the main group of SN Ia; this sub-group tends to be found in E or S0 galaxies. In \S6, we find the relationship between inverse B-band stretch factor and U, B, V, R and I-band luminosity, using colour-selected SNe Ia to avoid contamination from the outlying sub-group. We obtain a stretch - magnitude relationship for ``BV bluest'' SNe Ia without colour correction as $(2.28 \pm 0.42) \times s_{(B)}^{-1} - (21.49 \pm 0.41)$, which is steeper than the slope of the corresponding stretch-magnitude relation in ALT04. The r.m.s. of the residuals from this relationship is 0.17 mag. If we attempt to reduce the dispersion in luminosity by using colour corrections, we find that for distant (redshift $>$ 0.02) SNe Ia, $M_B = 0.98 \times s_{(B)}^{-1}- 2.28 \times (B - V)_{max} - 19.95$ with an r.m.s. of 0.27 mag. The r.m.s. becomes 0.12 mag when we select only SNe Ia hosted by E or S0 galaxies. In \S7, we statistically investigate the properties of dust in host galaxes, and find that the ratio of total to selective extinction $R$ can be consistent with that of dust in the Milky Way if we assume that SNe Ia have some diversity in their intrinsic colours.

Based on these results, in \S8, we discuss how to select subsets of SNe Ia which may be good distance indicators for cosmology. We find two possibilities: one is to use only ``BV bluest'' SNe Ia with broad light curves, and the other is to use only SNe Ia inside E or S0 galaxies. To use these samples for precision cosmology, we need to study the properties of SNe Ia from the low-z to the high-z universe.

\section*{Acknowledgments}

We are grateful to all the people who observed and analyzed the SNe Ia used in this paper. We are also grateful to members of the SDSS-II Supernova Survey and Supernova Cosmology Project for useful discussions. We thank Saul Perlmutter and David Rubin for discussion and comments on SNe Ia hosted by elliptical galaxies. We also thank Micheal Richmond for his carefully reading the manuscript. This work is supported in part by a JSPS core-to-core program ``International Research Network for Dark Energy'' and by a JSPS research grant (17104002).

\clearpage

\begin{figure}
\includegraphics[scale=0.45]{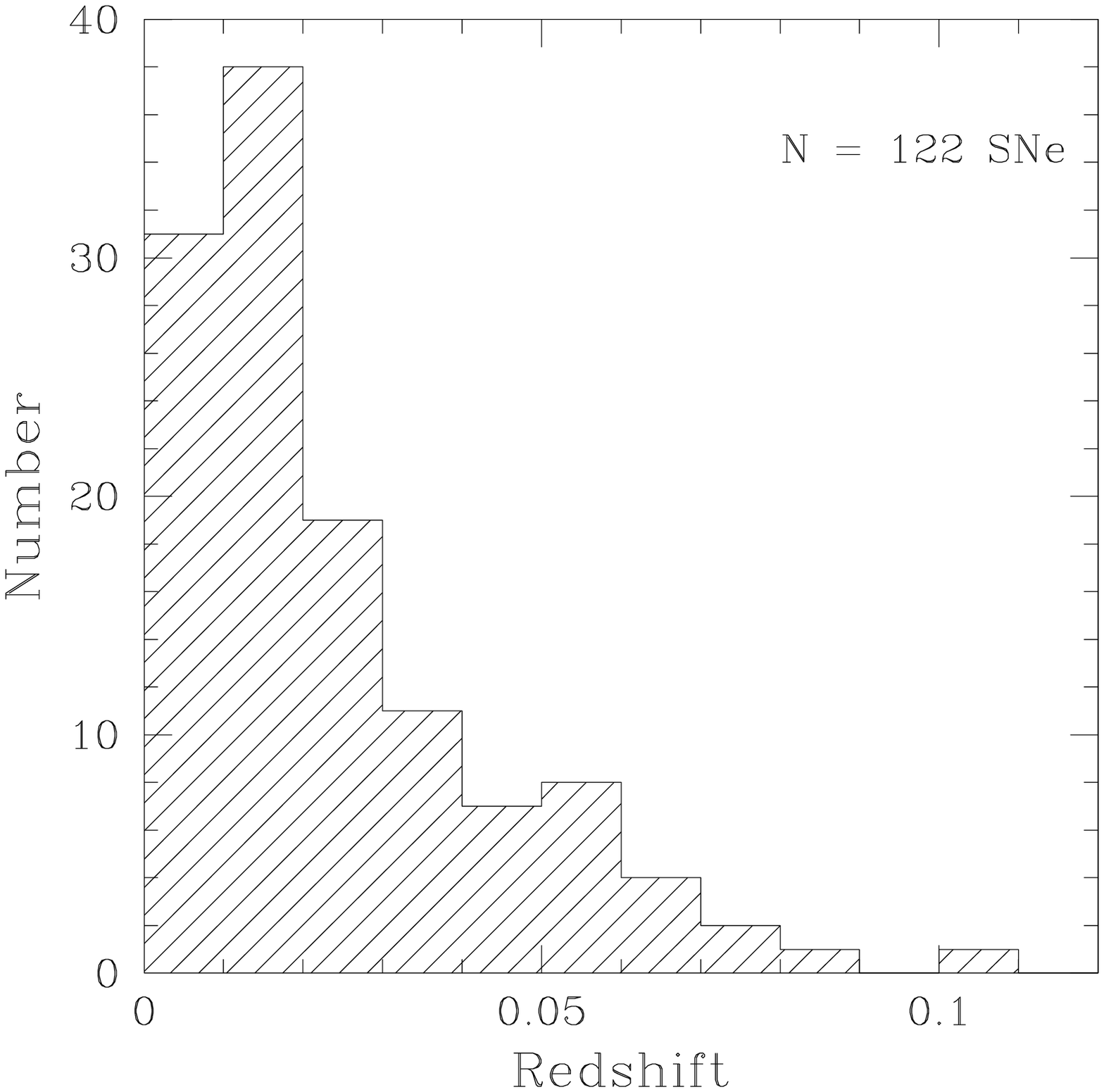}
\caption{Redshift distribution of SNe Ia used in this work. Most of them are at $z < 0.02$. \label{z_count}}
\end{figure}

\begin{figure}
\includegraphics[scale=0.45]{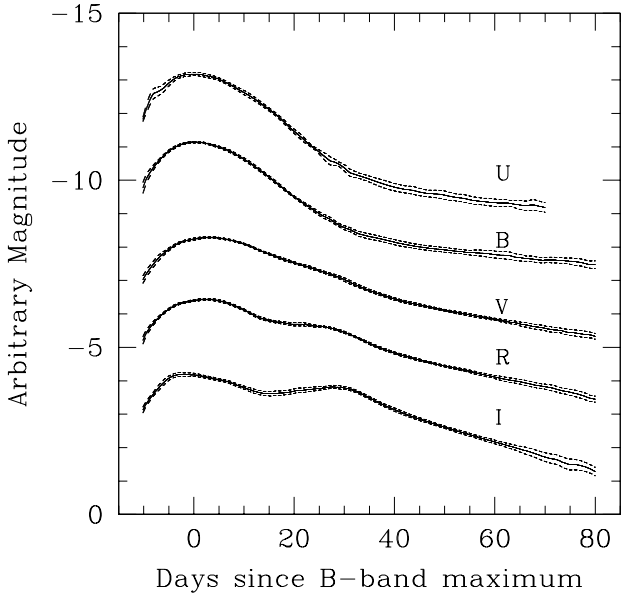}
\caption{U, B, V, R and I-band light curve templates (stretch factor = 1) with scatter. The r.m.s. (1$\sigma$) are derived from residuals of composited light curves and shown in dashed line. The B, V, R, I-band templates are from -10 days to +80 days, and the U-band template is from -10 days to 70 days after B-band maximum. The U-band template has larger dispersion than B, V, R and I-band templates because U-band photometry is poorer than B, V R and I-band photometries. \label{templates}}
\end{figure}

\begin{figure}
\includegraphics[scale=0.45]{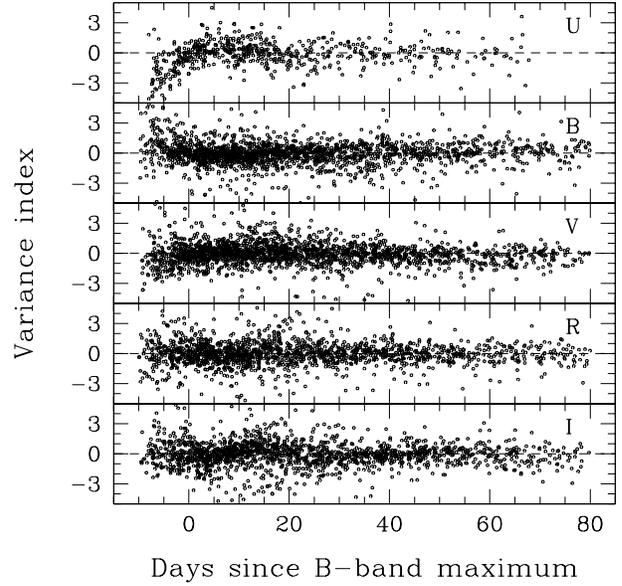}
\caption{Residuals from light curve template fitting in U, B, V, R and I-band magnitude. All epochs of 122 SNe Ia are plotted after fitting to the templates. Variance index is defined as $\Delta m_{X} / \sqrt{\sigma^2_{template} + \sigma^2_{photometry}}$, here, $\Delta m_X$ is the residual between the X-band template and the observation (plus sign means that photometry is fainter than template). $\sigma_{template}$ and $\sigma_{photometry}$ are typically in the order of 0.01 mag. Early phase U-band photometry (day $<$ 0) are systematically brighter than the template. A possible reason of the excess is K-correction. \label{day_chi}}
\end{figure}

\begin{figure}
\includegraphics[scale=0.45]{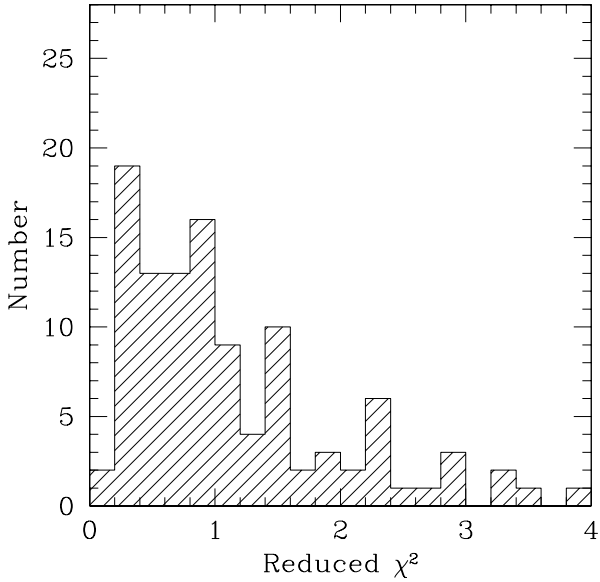}
\caption{Distribution of reduced $\chi^2$ of the fitting. Most of SNe Ia well fit in the templates, but there are some SNe Ia not well fitted in the templates. {\bf $\chi^2 / d.o.f. > 4$ :} SN1994T, SN1996bo, SN1997cn, SN1998aq, SN1998de, SN1999ao, SN1999by, SN1999da, SN2002cx, SN2003du. {\bf $\chi^2 / d.o.f. > 3$ :} SN1992A, SN1992ag, SN1998bp, SN1999aw. A few of them are known as spectroscopically peculiar SNe Ia, but some of the others are known as "Branch normal" SNe Ia. \label{chi_count}}
\end{figure}

\begin{figure}
\includegraphics[scale=0.45]{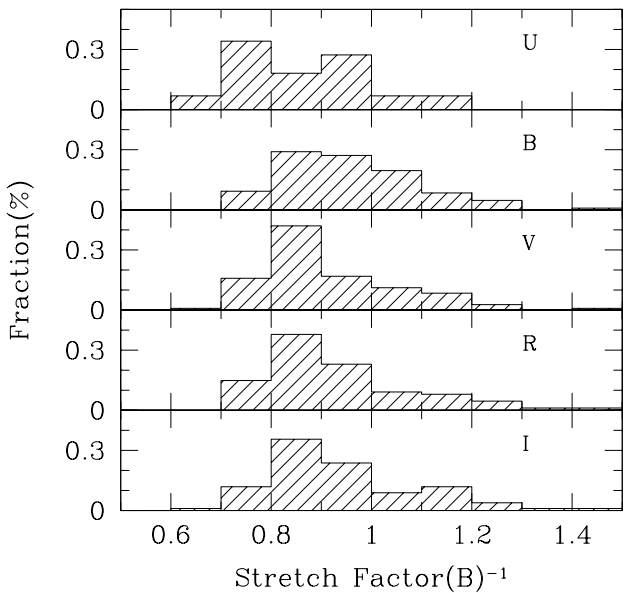}
\caption{Histograms of the stretch factors in U, B, V, R and I-band. There are 44 SNe Ia in U-band, 108 SNe Ia in B-band, 108 SNe Ia in V-band, 88 SNe Ia in R-band, 102 SNe Ia in I-band. \label{sf_count}}
\end{figure}

\begin{figure}
\begin{tabular}{cc}
\begin{minipage}{0.5\hsize}
\begin{center}
\includegraphics[scale=0.23]{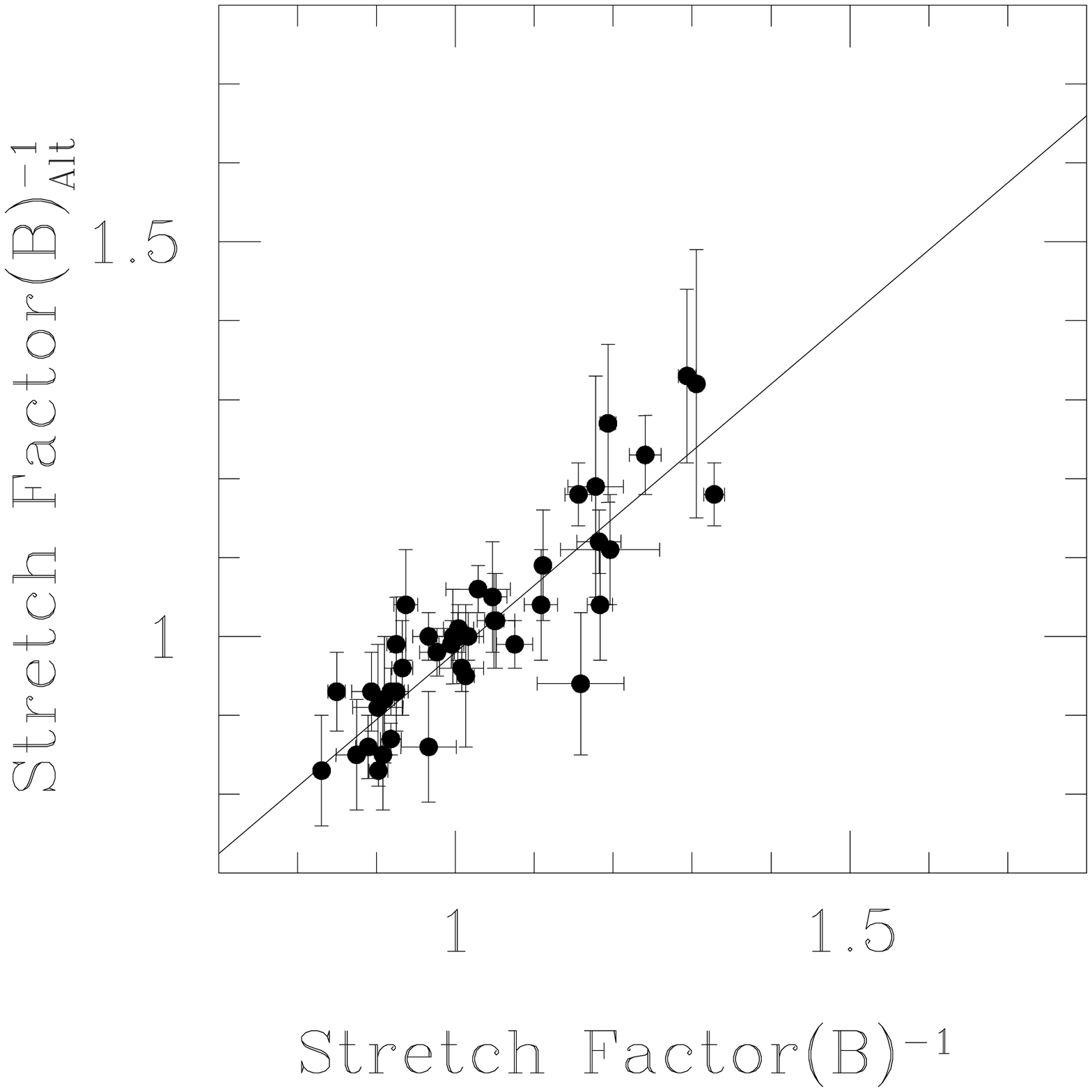}
\end{center}
\end{minipage}
\begin{minipage}{0.5\hsize}
\begin{center}
\includegraphics[scale=0.23]{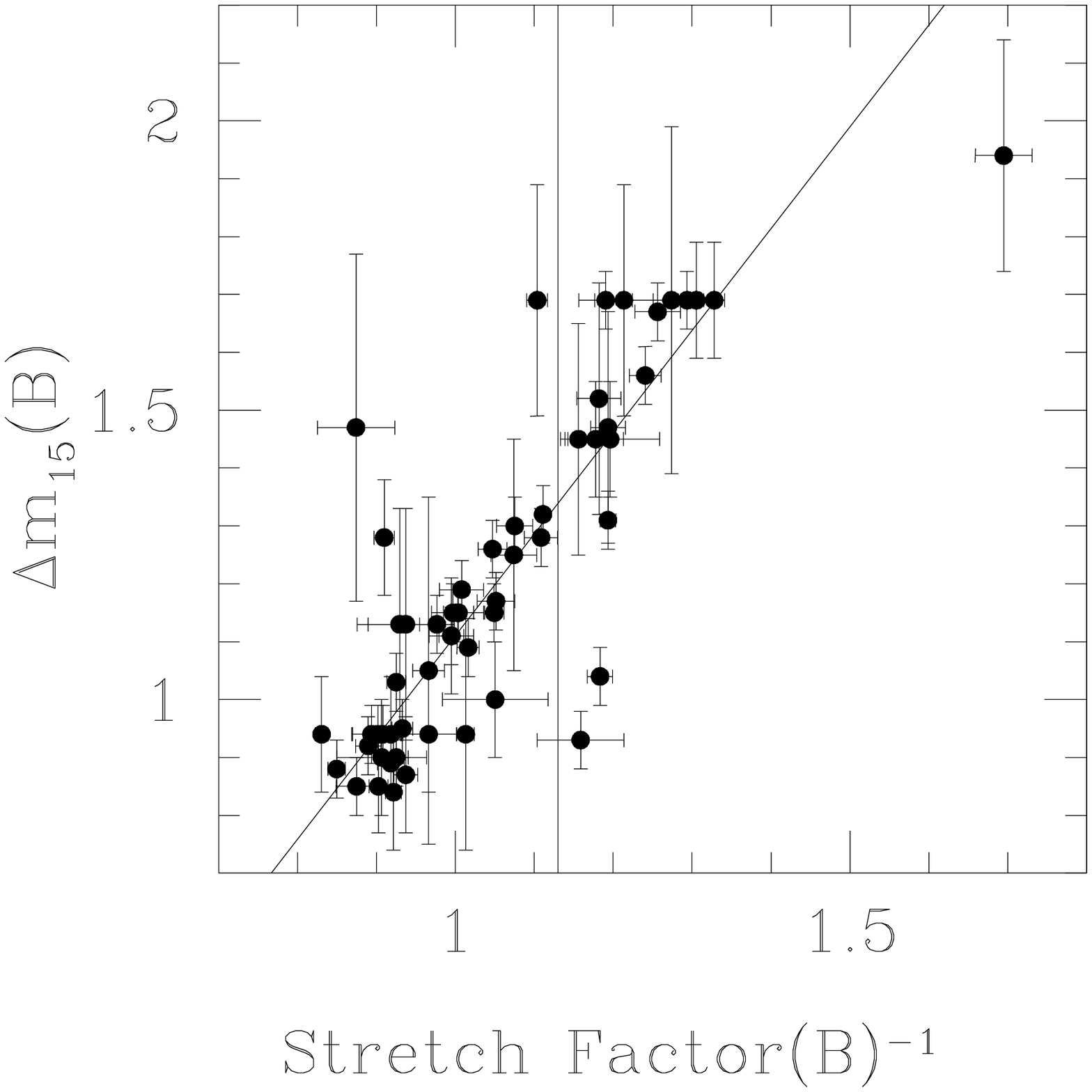}
\end{center}
\end{minipage}
\end{tabular}
\caption{Comparison with ALT04. The left panel shows a relation between inverse B-band stretch factors. The right panel shows the relation between inverse B-band stretch factor (this work) and $\Delta m_{15}$ from ALT04. The solid line is a regression line derived from least $\chi^2$ fitting. \label{Bsf_asf_dm15}}
\end{figure}

\begin{figure}
\includegraphics[scale=0.45]{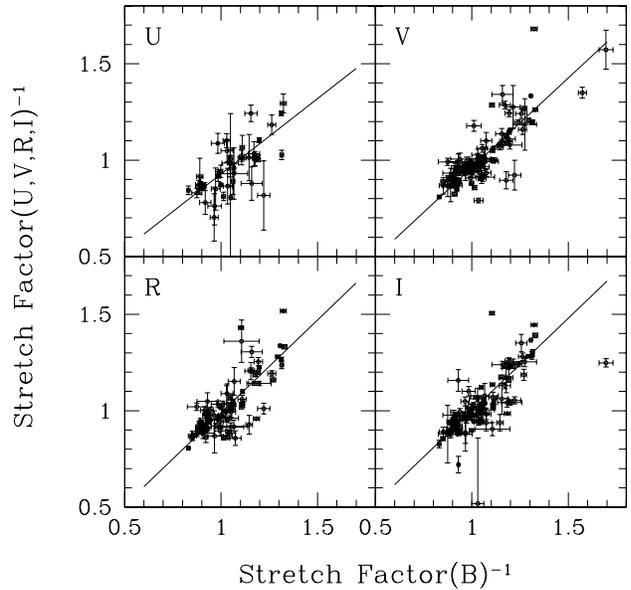}
\caption{Relations between the inverse B-band stretch factor and inverse U, V, R, I-band stretch factors of all SNe Ia. The solid lines are regression lines (see also Table\ref{Bsf_sf_tab}). \label{Bsf_sf}}
\end{figure}

\begin{figure}
\includegraphics[scale=0.45]{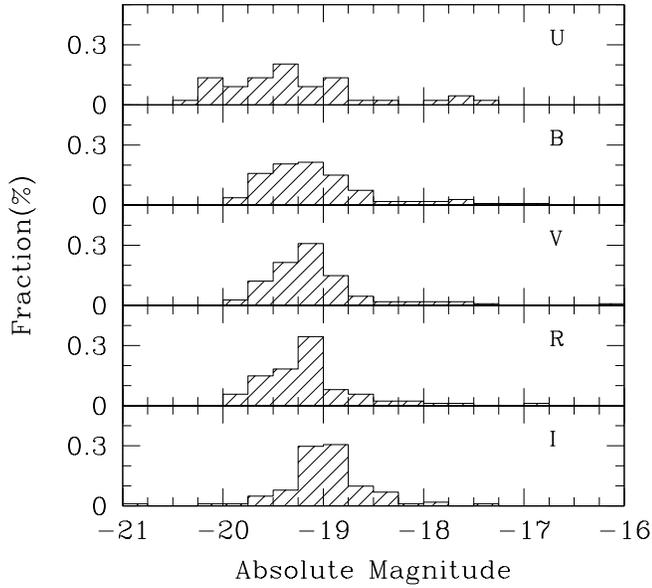}
\caption{Histograms of absolute magnitudes in U, B, V, R and I-band. The magnitudes are corrected for Galactic dust extinction, but not for host galaxy dust extinction. Bluer bands such as U and B-band have broader distributions than redder bands. \label{mag_count}}
\end{figure}

\begin{figure}
\includegraphics[scale=0.45]{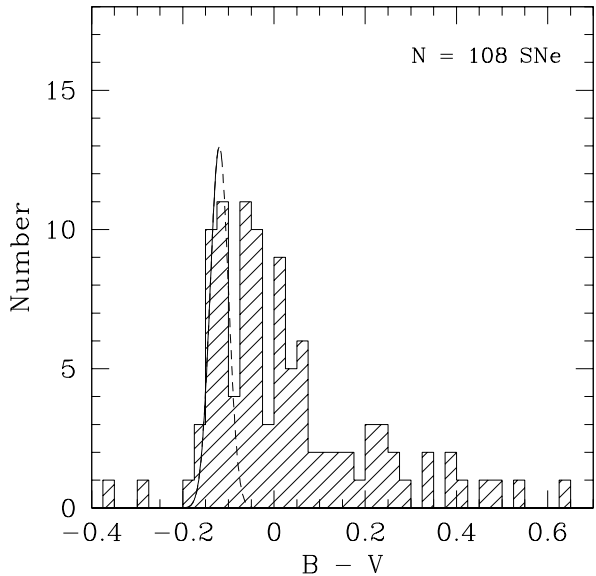}
\caption{Rest frame distribution of $(B-V)_{max}$ without host galaxy dust correction. Tail in redder side can be attributed host galaxy dust reddening. We can regard bluer SNe Ia may be less affected by host galaxy dust. Median $(B-V)_{max}$ of one-side Gaussian fit using only bluest SNe Ia is -0.12 +/- 0.02. This value is consistent with ALT04.\label{BV_count}}
\end{figure}

\begin{figure}
\includegraphics[scale=0.45]{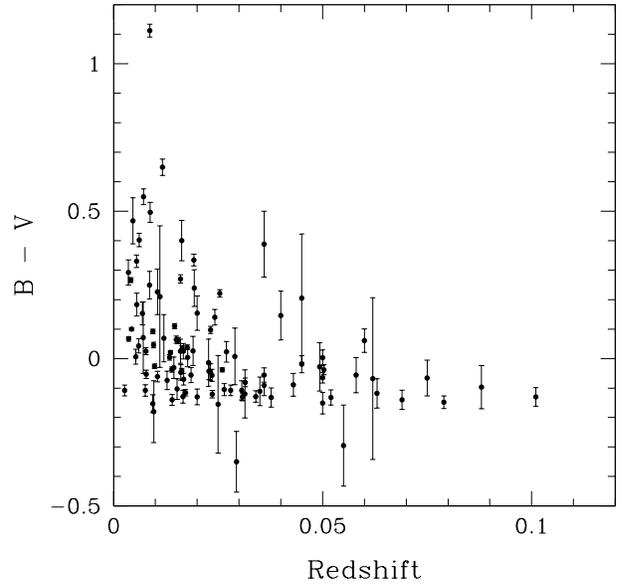}
\caption{$(B-V)_{max}$ distribution as a function of redshift. Error bars include only photometric errors and fitting errors (peculiar velocity of galaxies are not included). Redder SNe Ia tend to be missing at larger redshift because they are fainter than bluer SNe Ia which expected to be free from host galaxy dust. Also this figure shows there are no apparent trend in the bluest end of $(B-V)_{max}$ colour along redshift at the range of $z \lesssim 0.1$.\label{z_BV}}
\end{figure}

\begin{figure}
\includegraphics[scale=0.45]{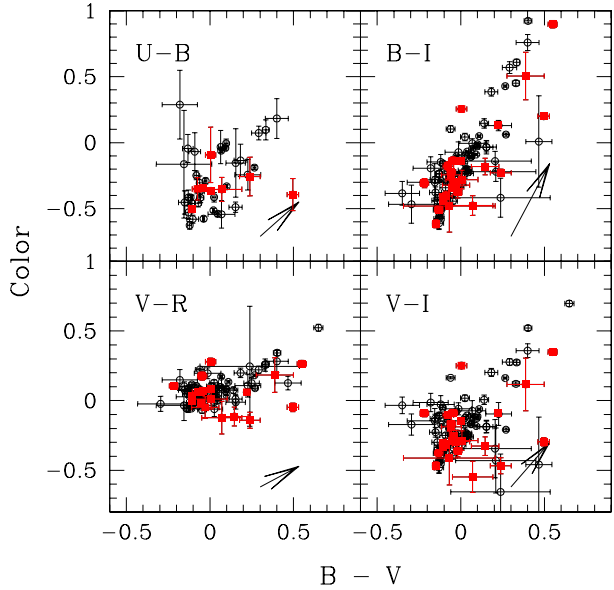}
\caption{Colour-colour diagram of SNe Ia at B-band maximum. Filled red squares are SNe Ia hosted by E or S0 galaxies. Arrows show direction and size of reddening when we assume that absorption properties of host galaxies dust are same as Galactic dust ($R_{U} = 5.434, R_{B} = 4.315, R_{V} = 3.315, R_{R} = 2.673, R_{I} = 1.940$ from \protect\citealt{sch98}) and $A_{B} = 1.0 mag$. Even if we selected only SNe Ia hosted by Elliptical galaxies, they are plotted along the Galactic dust reddening direction. The colour have been corrected for Galactic dust reddening, but not for host galaxy dust reddening. \label{BV_colour}}
\end{figure}

\begin{figure}
\includegraphics[scale=0.45]{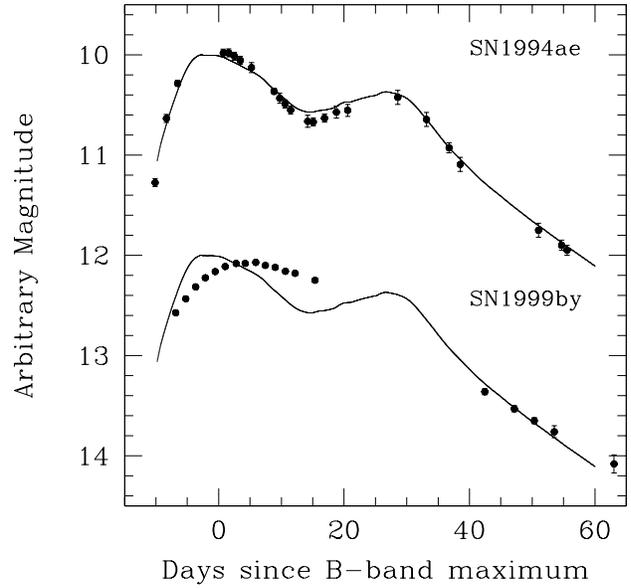}
\caption{Examples of normal I-band light curve and peculiar I-band light curve. SN1994ae is a spectroscopically normal SN Ia, and there is a "bump" in the I-band light curve. SN1999by is a spectroscopically peculiar SN Ia, and there is no "bump" in the I-band light curve. \label{IbandLC}}
\end{figure}

\begin{figure}
\includegraphics[scale=0.45]{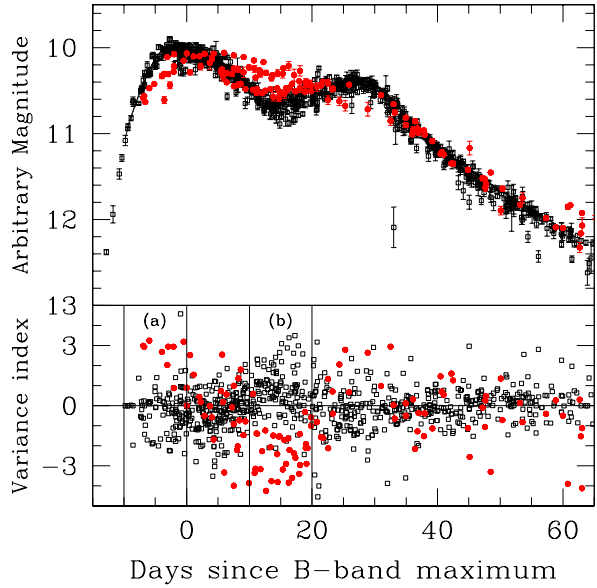}
\caption{I-band light curves of 71 selected SNe Ia (top panel) and distribution of the epochs (bottom panel). Variance index is defined as $\Delta m_I/\sqrt{\sigma^2_{template} + \sigma^2_{photometry}}$, where $\Delta m_I$ is the residual between the I-band template and the observation, $\sigma_{photometry}$ is the photometric error and $\sigma_{template}$ is the template error. We classified the SNe into two types, normal I-band light curves (open square) and peculiar I-band light curves (filled red circle), according to photometry points (see text). \label{Ibandclass}}
\end{figure}

\clearpage
\begin{figure}
\includegraphics[scale=0.45]{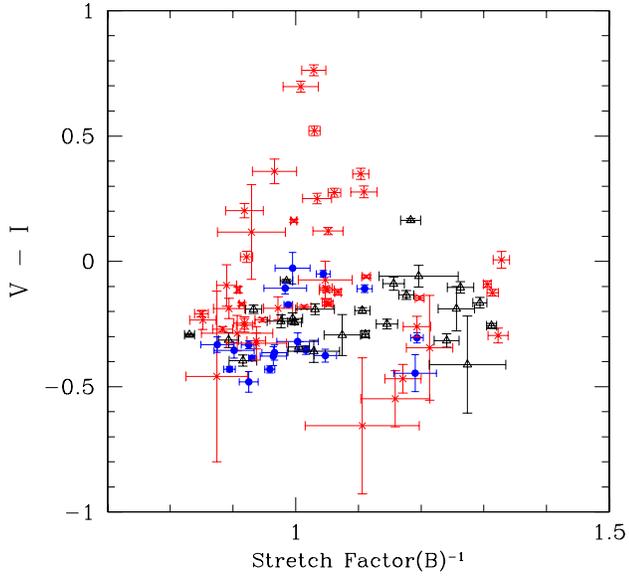}
\caption{Distribution of $(V-I)_{max}$ as a function of the inverse B-band stretch factor. Filled blue circles are "BV bluest" SNe Ia, red crosses are "BV redder" SNe Ia and open triangles are the other SNe Ia. \label{Bsf_VI}}
\end{figure}

\begin{figure}
\includegraphics[scale=0.45]{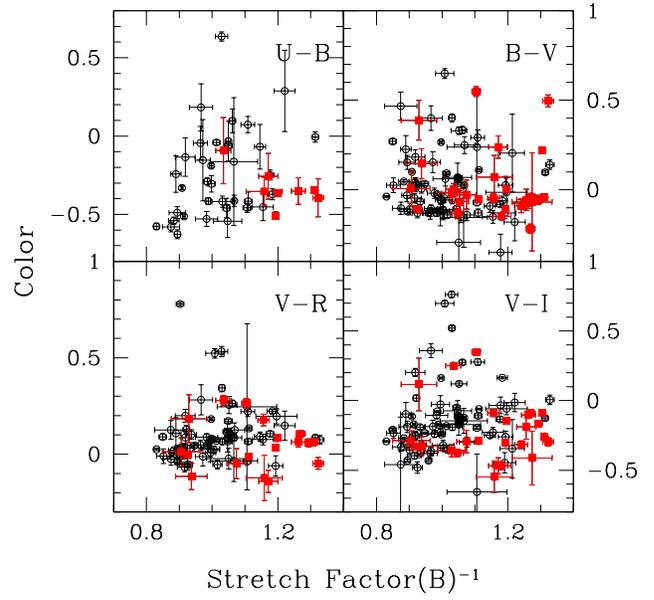}
\caption{Distribution of rest frame colours as a function of the inverse B-band stretch factor. Filled red squares are SNe hosted by E or S0 galaxies. The colour have been corrected for Galactic dust reddening, but not for host galaxy dust reddening. \label{Bsf_colour}}
\end{figure}

\begin{figure}
\includegraphics[scale=0.45]{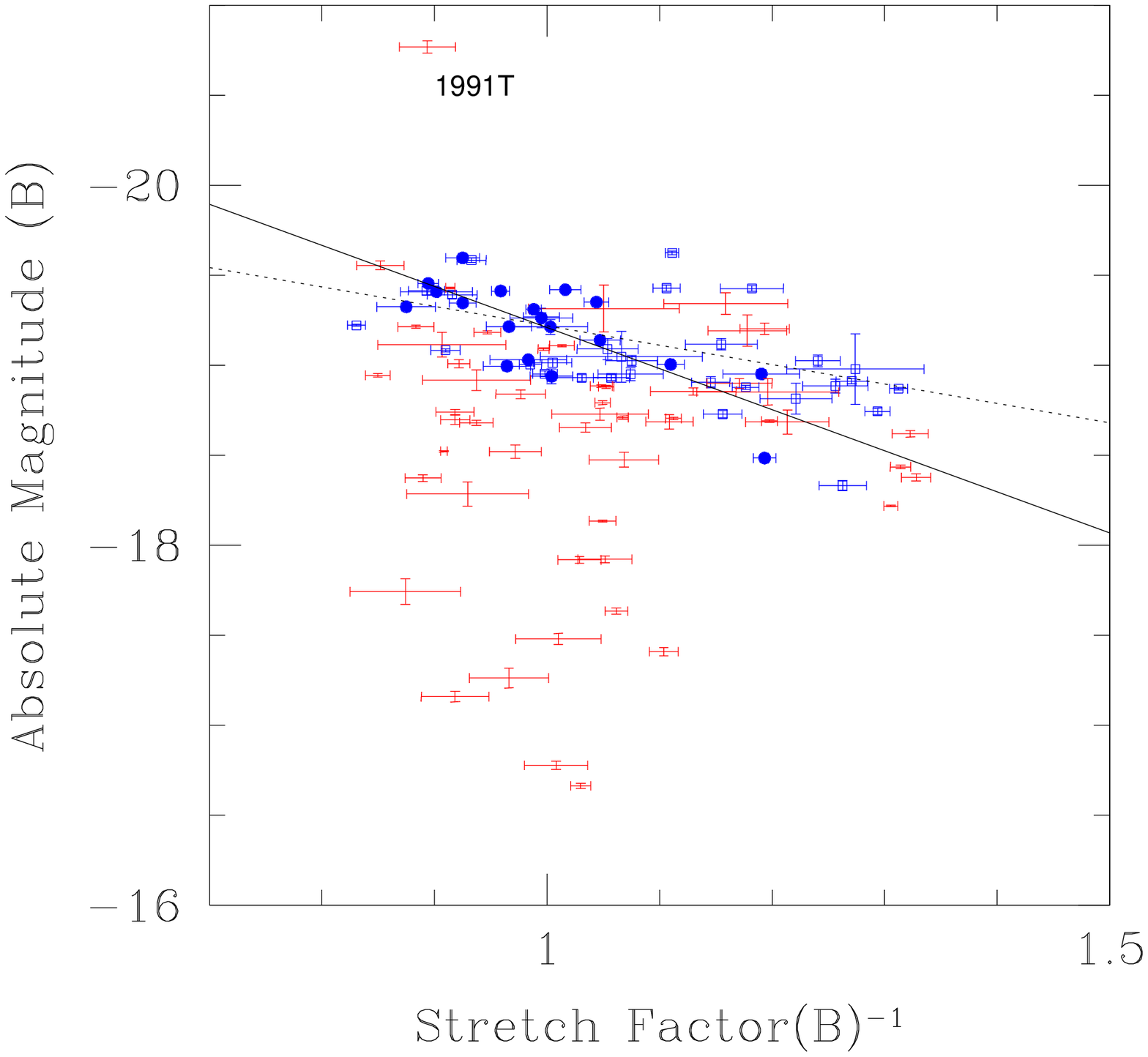}
\caption{Relations between inverse B-band stretch factor and absolute B-band magnitude. Filled blue circles are "BV bluest" SNe Ia, open blue squares are "BV bluer" SNe Ia and the other symbols are for other SNe Ia. The solid line is a relation between inverse B-band stretch factor and B-band magnitude derived from "BV bluest" SNe Ia, while the dashed line is a relation derived from "BV bluest" + "BV bluer" SNe Ia. Redder and fainter SNe Ia than those "BV bluest" or "BV bluer" SNe Ia may be affected by host galaxy dust. \label{Bsf_Bmax}}
\end{figure}

\begin{figure}
\includegraphics[scale=0.45]{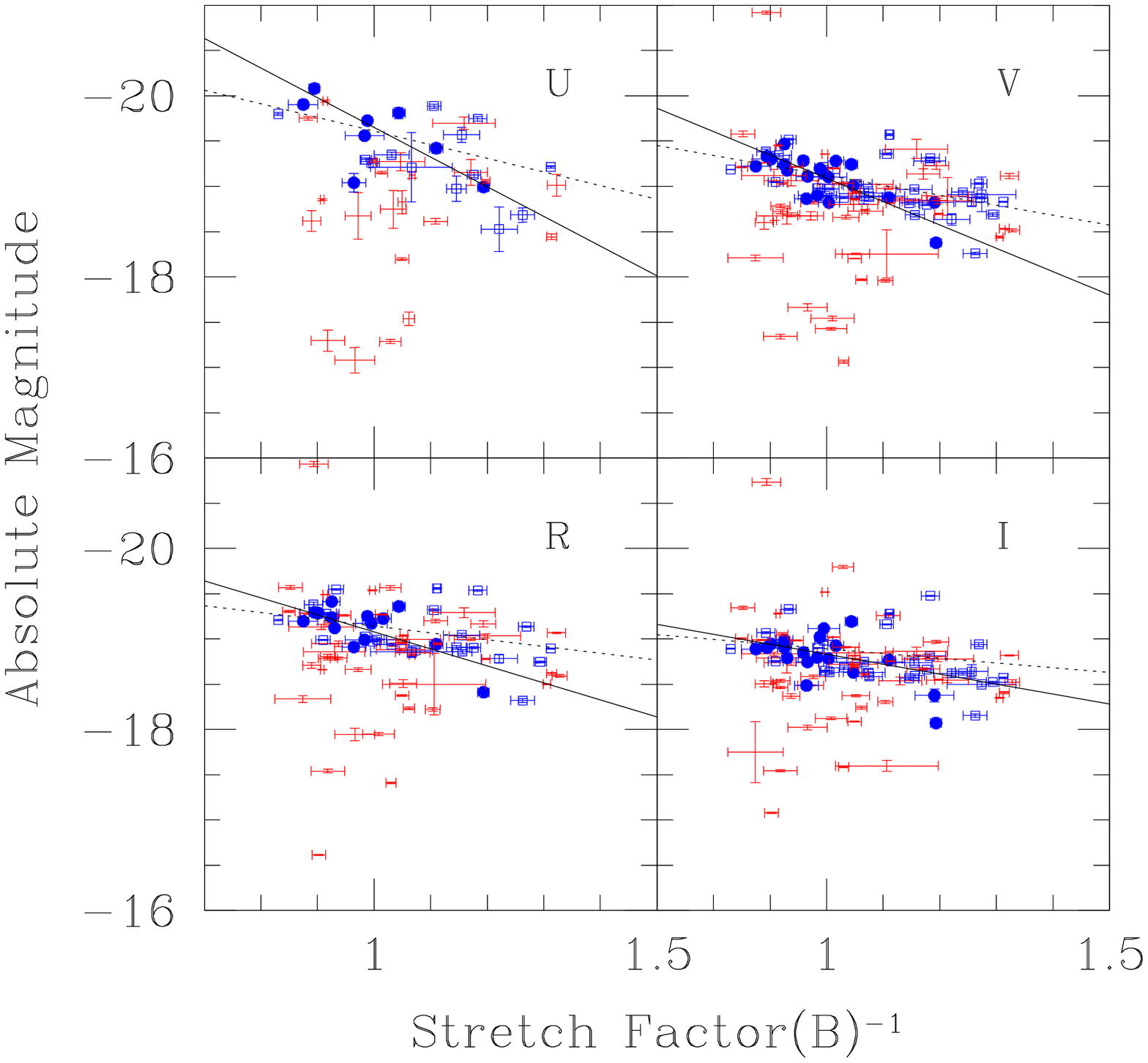}
\caption{Relations between inverse B-band stretch factor and absolute magnitude of U, V, R and I-band. Filled blue circles are "BV bluest" SN Ia, open blue squares are "BV bluer" SNe Ia and the other symbols are for other SNe Ia. The solid line is a relation between inverse B-band stretch factor and B-band magnitude derived from "BV bluest" SNe Ia, while the dashed line is a relation derived from "BV bluest" + "BV bluer" SNe Ia. \label{Bsf_UVRImax}}
\end{figure}

\begin{figure}
\includegraphics[scale=0.45]{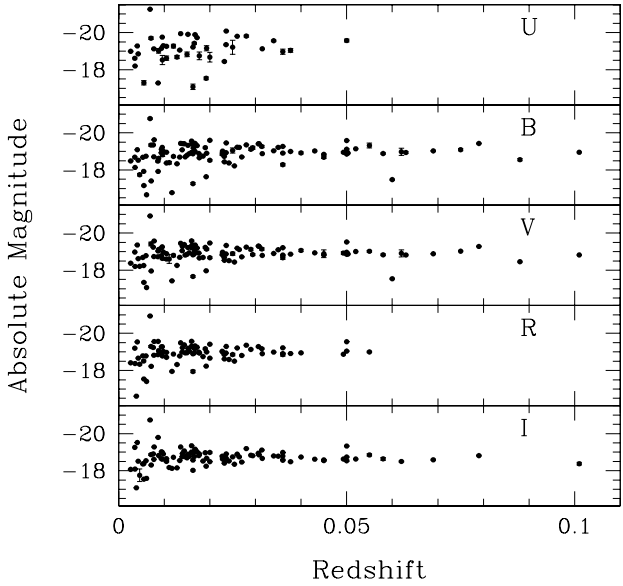}
\caption{Absolute magnitude distribution as a function of redshift. Magnitude is corrected for Galactic dust extinction, but not for host galaxy dust extinction. The bright outlier near z = 0.007 is SN1991T. If peculiar velocity is not negligible, the number of brighter SNe Ia than the average increase in lower-z (also fainter SNe Ia increase, but we can't distinguish them from the SNe Ia which extincted by host galaxy dust). However, there is no considerable effect caused by peculiar velocity. See also Table \ref{Bmag_redshift}. \label{z_max}}
\end{figure}

\begin{figure}
\includegraphics[scale=0.45]{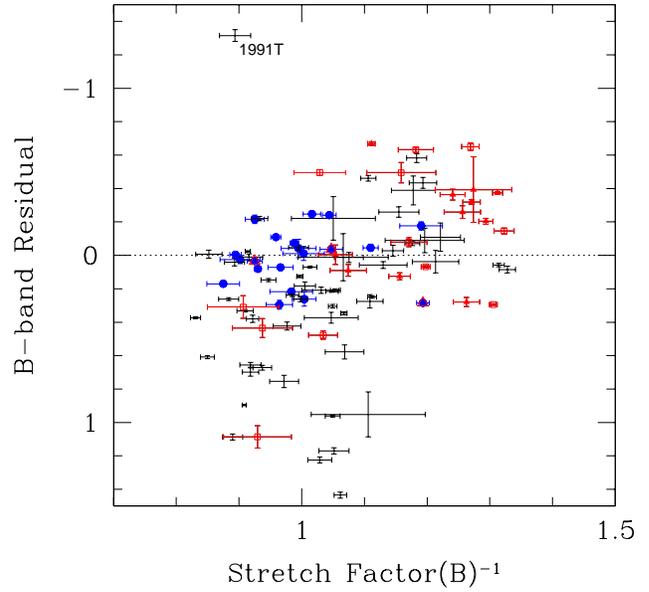}
\caption{Residuals from the stretch - magnitude relation derived from "BV bluest" SNe Ia without dust/colour correction. Filled blue circles are "BV bluest" SNe Ia (sample 7A in Table \ref{Bsf_Bmag_type}), red symbols are SNe Ia hosted by E or S0 galaxies (sample 7B in Table \ref{Bsf_Bmag_type}), filled red triangles are "BV bluer" SNe Ia hosted by E or S0 galaxies (sample 7C in Table \ref{Bsf_Bmag_type}), open red triangles with filled blue circle are "BV bluest" SNe Ia hosted by E or S0 galaxies (sample 7D in Table \ref{Bsf_Bmag_type}) and the other symbols are for other SNe Ia. \label{Bsf_Bmag_ES0}}
\end{figure}

\begin{figure}
\includegraphics[scale=0.45]{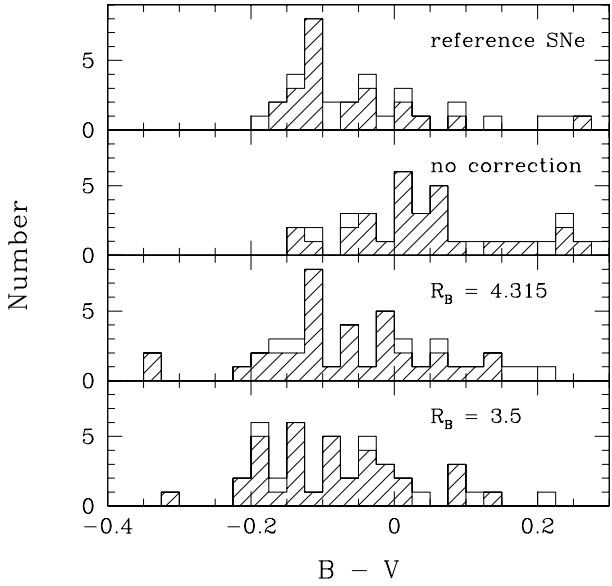}
\caption{Distributions of (B-V)$_{max}$ of dust free and dust corrected SNe Ia using B-band brightness. The top panel is reference SNe Ia which are mostly dust free. The second panel is for SNe Ia which are significantly affected by host galaxy dust, but without dust corrections. The third panel is after dust correction using $R_{B} = 4.315$ (\protect\citealt{sch98}), and the bottom panel is dust correction with $R_{B} = 3.5$ (ALT04). Shaded bars show type Ia  SNe with broader light curve shape, $s_{(B)}^{-1} < 1.1$, which are expected to be spectroscopically "normal" SNe Ia (see Figure \ref{Bsf_VI}). \label{BV_count_correct_B}}
\end{figure}

\begin{figure}
\includegraphics[scale=0.45]{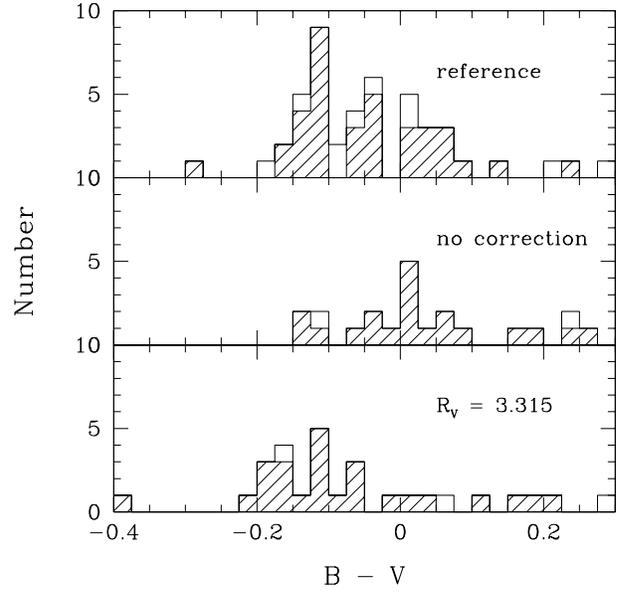}
\caption{Distribution of (B-V)$_{max}$ of dust free and dust corrected SNe Ia using V-band brightness. The top panel is reference SNe Ia which are mostly dust free. The middle panel is for SNe Ia which are significantly affected by dust, but without dust corrections. The bottom panel is after dust correction with $R_{V} = 3.315$ (\protect\citealt{sch98}). Shaded bars show SNe Ia with broader light curve shape, $s_{(B)}^{-1} < 1.1$, which are expected to be spectroscopically "normal" SNe Ia (see Figure \ref{Bsf_VI}). \label{BV_count_correct_V}}
\end{figure}
\bsp

\label{lastpage}

\end{document}